\documentclass[aps,prc,twocolumn,superscriptaddress]{revtex4-2}
\usepackage{amssymb}
\usepackage{amsmath}
\usepackage{newtxtext}
\usepackage[upint]{newtxmath}
\usepackage{mathtools}
\usepackage{bm}
\usepackage{graphicx}
\usepackage{hyperref}
\usepackage[all]{hypcap}

\pdfoutput=1

\newcommand{\const}{\operatorname{const}}

\begin{document}

\title{Loss rate of ultracold neutrons due to the absorption by trap walls
in large material traps}
\author{Pavel\,D.~Grigoriev}
\email[Corresponding author, e-mail:]{grigorev@itp.ac.ru}
\affiliation{L.\,D.~Landau Institute for Theoretical Physics, 142432, Chernogolovka, Russia}
\affiliation{Theoretical Physics and Quantum Technologies Department, National University of 
Science and Technology ''MISIS'', 119049, Moscow, Russia} 
\affiliation{National Research University Higher School of Economics, Moscow, 101000, Russia}

\author{Vladislav\,D.~Kochev}
\affiliation{Theoretical Physics and Quantum Technologies Department, National University of 
Science and Technology ''MISIS'', 119049, Moscow, Russia}
\affiliation{%Department of Condensed Matter, 
	NRC Kurchatov Institute, 123182, Moscow, Russia}
\affiliation{L.\,D.~Landau Institute for Theoretical Physics, 142432, Chernogolovka, Russia}

\author{Victor A. Tsyplukhin}
\affiliation{Theoretical Physics and Quantum Technologies Department, National University of 
Science and Technology ''MISIS'', 119049, Moscow, Russia}

\author{Alexander\,M.~Dyugaev}
\affiliation{L.\,D.~Landau Institute for Theoretical Physics, 142432, Chernogolovka, Russia}

\author{Ilya\,Ya.~Polishchuk}
\affiliation{%Department of Condensed Matter, 
	NRC Kurchatov Institute, 123182, Moscow, Russia}
\affiliation{Theoretical Physics Department, Moscow Institute For Physics and Technology, 141700, 
Dolgoprudnii, Russia}

\date{\today}

\begin{abstract}
The most accurate neutron lifetime measurements now use the material or
magnetic traps of ultracold neutrons (UCN). The precision of these
experiments is determined by the accuracy of estimating the neutron loss
rate. In material UCN traps the main source of neutron losses is the
absorption by trap walls. In this paper we analyze the standard methods and
their approximations for the calculation of UCN absorption rate by the walls
of material traps. We emphasize the approximations used both in the standard
analytical formulas and in the numerical Monte-Carlo simulations. For the
two simplest trap geometries, rectangular and cylindrical, we obtain
analytical formulas for this absorption rate provided the UCN velocity 
distribution is isotropic at trap bottom. 
Then we perform numerical calculations of UCN velocity distribution and absorption 
rate taking into account  the diffuse elastic UCN reflections by trap walls 
obeying two different laws: Lambert's cosine law and isotropic reflection.  
We compare the results with the standard estimation methods and discuss the differences. 
We indicate the difference between the UCN \emph{number} and \emph{density} velocity distribution. 
%The difference turned out to be considerable and especially important for the size extrapolation procedure, always used in the standard estimates of UCN losses. 
Our results may be useful to resolve
the puzzling four-second discrepancy between the magnetic and material-trap
measurements of neutron lifetime.
\end{abstract}

\maketitle

\section{Introduction}

The neutron $\upbeta$-decay $n \to p+e^{-}+\bar{\nu}_e$ plays an important
role in cosmology, astrophysics and elementary particle physics (see \cite%
{Dubbers/2011,Abele/2008,GONZALEZALONSO2019165,Universe2023,MuSolf/2008,WietfeldtColloquiumRMP2011,Abele2023}
for reviews). The primordial light element abundance is very sensitive to
the exact value of neutron mean lifetime $\tau_\text{n}$ \cite%
{Dubbers/2011,Mathews2005}. The precise $\tau_\text{n}$-measurements,
combined with decay correlations in polarized-neutron decay experiments \cite%
{Beam2019PhysRevLett.122.242501,Beck2024,PhysRevLett.105.181803,PhysRevC.101.035503}%
, test the standard model and give the coupling constants of the weak
interaction \cite%
{Dubbers/2011,Abele/2008,GONZALEZALONSO2019165,PhysRevLett.105.181803,Beam2019PhysRevLett.122.242501,
PhysRevC.101.035503,Beck2024,Ivanov2013,Universe2023}%
. 
%precise measurements of neutron lifetime $\tau_\text{n}$ are important for
The measurements of neutron electric dipole moment (EDM) \cite%
{AbelEDM2020,Pospelov2005,Baker/2006,SerebrovJETPLetters2014} impose the
upper limits on CP violation. The resonant transitions between discrete
quantum energy levels of neutrons in the earth gravitational field \cite%
{NesvizhevskyNature2002,UCNResonancePhysRevLett.112.151105,Jenke2011} probe
the gravitational field on a micron length scale and impose constraints on
dark matter.

These and other neutron experiments mostly employ the ultracold neutrons
(UCN) with energy $E$ lower than either the neutron optical potential of
typical materials or the Zeeman energy of neutron spin in available magnetic
fields, i.e. $E\lesssim$ 300~neV \cite%
{Golub/1991,Ignatovich/1990,Ignatovich1996,PhysRevLett.105.181803,PhysRevC.101.035503,AbelEDM2020,
Pospelov2005,Baker/2006,SerebrovJETPLetters2014,NesvizhevskyNature2002,UCNResonancePhysRevLett.112.151105,
Jenke2011,Serebrov2008PhysRevC.78.035505,ArzumanovPhysLettB2015,Serebrov2017,Serebrov2018PhysRevC.97.055503,
Review2019Pattie,MAMBOPhysRevLett.63.593,PICHLMAIER2010,Huffman2000,PhysRevC.94.045502,PhysRevC.95.035502,
Ezhov2018,Pattie2018,Gonzalez2021}%
. In material traps UCN can be trapped for many minutes in specially
designed ''neutron bottles'' \cite%
{Serebrov2008PhysRevC.78.035505,ArzumanovPhysLettB2015,Serebrov2017,Serebrov2018PhysRevC.97.055503,
Review2019Pattie,MAMBOPhysRevLett.63.593,PICHLMAIER2010}%
, where the earth gravitational field 100 neV per meter plays an important
role in UCN storage and manipulation \cite%
{Golub/1991,Ignatovich/1990,Ignatovich1996,Serebrov2008PhysRevC.78.035505,ArzumanovPhysLettB2015,
Serebrov2017,Serebrov2018PhysRevC.97.055503,Review2019Pattie}%
. The Fomblin grease is currently used to cover the UCN trap walls \cite%
{Serebrov2008PhysRevC.78.035505,ArzumanovPhysLettB2015,Serebrov2017,Serebrov2018PhysRevC.97.055503,
Review2019Pattie,MAMBOPhysRevLett.63.593,PICHLMAIER2010}
in the bottle UCN experiments and allows reaching the very high accuracy of
neutron lifetime measurements in large gravitational traps \cite%
{Serebrov2018PhysRevC.97.055503}: $\tau _\text{n}$ = 881.5 $\pm$ 0.7 (stat) $%
\pm$ ~0.6 (syst) s. The neutron magnetic moment of 60 neV/T allows
magneto-gravitational trapping of UCN \cite%
{Huffman2000,PhysRevC.94.045502,PhysRevC.95.035502,Ezhov2018,Pattie2018,Gonzalez2021}%
, giving even higher claimed accuracy of latest UCN-$\tau$ measurements \cite%
{Gonzalez2021}: $\tau _\text{n} \approx $ 877.7 $\pm $ 0.28 (stat) $%
_{-0.16}^{+0.22}$ (syst) s. The neutron lifetime measured using magnetic
traps \cite{Gonzalez2021} is about 4 seconds smaller than in the UCN
material-bottle experiments \cite{Serebrov2018PhysRevC.97.055503}, which is
beyond the $3\sigma$ deviation.

The most precise time-of-flight $\tau _\text{n}$-measurements with the beam
of cold neutrons give $\tau _\text{n}$ = 887.7 $\pm $ 1.2 (stat) $\pm $ 1.9
(syst) s \cite%
{BeamPhysRevC.71.055502,BeamPhysRevLett.111.222501,BeamReview2020}, which is
about 10 seconds greater than in the UCN magnetic traps. The large
discrepancy between $\tau _\text{n}$ measured using the beam and UCN
material- or magnetic-trap methods is called the ''neutron lifetime puzzle''
and receives extensive discussion till now \cite%
{BeamReview2020,DarkMatter2021PhysRevD.103.035014,Serebrov2021PhysRevD.103.074010,Serebrov2019,Koch2024}%
. Presumably, this is due to systematic errors in beam experiments \cite%
{Serebrov2021PhysRevD.103.074010}, but unaccounted UCN losses in the bottle
material and magnetic trap $\tau _\text{n}$ measurements have not yet been
excluded. The analysis of neutron $\upbeta$-decay asymmetry \cite%
{Dubbers2019} suggests that this discrepancy is unlikely caused by new
physics, for example, by dark matter and additional neutron decay channels
or excited states \cite{DarkMatter2021PhysRevD.103.035014,BeamReview2020}.

The precision of $\tau _\text{n}$ measurements using UCN traps, both
material and magnetic, is determined by the accuracy of estimating the
neutron loss rate from the traps, which is the main source of systematic
errors \cite%
{Golub/1991,Ignatovich/1990,Ignatovich1996,Goremychkin2017,Serebrov2008PhysRevC.78.035505,
ArzumanovPhysLettB2015,Serebrov2017,Serebrov2018PhysRevC.97.055503,Review2019Pattie,Ezhov2018,
Pattie2018,Gonzalez2021}%
. The main UCN loss mechanism from magnetic traps, as listed in Table II of
Ref. \cite{Gonzalez2021}, include (i) UCN spin depolarization, for example,
because of the nonuniform magnetic field and, hence, its nonzero
perpendicular-to-spin component; (ii) heated UCNs; (iii) residual gas
scattering; (iv) uncleaned higher-energy UCNs. In this paper we do not
discuss the accuracy of magnetic-trap UCN experiments and consider only the
material traps.

The material UCN traps are, usually, coated with Fomblin grease, providing
the highest accuracy of $\tau _\text{n}$ measurements. The Fomblin grease
has the optical potential barrier $V_{0}^{\text{F}}\approx $ 106~neV. The
probability of neutron absorption by such a wall is $\sim 10^{-5}$ per
collision \cite%
{Golub/1991,Ignatovich/1990,Ignatovich1996,WietfeldtColloquiumRMP2011}. In
addition, the inelastic UCN scattering by trap walls, when the neutron
absorbs a thermal excitation, increases the UCN energy above the potential
barrier $V_0$ and also leads to neutron losses. Therefore, the neutron
lifetime $\tau _\text{n}$ is estimated by the double extrapolation of the
measured lifetime $\tau_\text{m}<\tau_\text{n}$ of UCN stored in the trap to
zero temperature (thermal extrapolation) and to an infinite trap size
(geometrical or size extrapolation). The extrapolation interval is rather
large, usually, $\tau _\text{n}-\tau_\text{m}\gtrsim 20$ s. This limits the
precision of $\tau _\text{n}$ measurements in material traps, because the
estimate of UCN loss rate with an accuracy better than 5\% is very
complicated.

The UCN absorption probability depends on the angle of incidence during each
collision and, hence, on UCN height and angular velocity distribution. 
The usually applied assumption \cite{Serebrov2018PhysRevC.97.055503} 
of the uniform distribution of neutron velocity direction with respect 
to the trap surface at any height may be violated 
%for the collisions with side walls 
because the vertical UCN velocity component of each neutron during its motion
depends on the height above trap bottom due to gravity. 
%Below we analyze this and some other assumptions. 
This difficulty can
be overcome by Monte-Carlo simulations \cite%
{Fomin2023,Fomin2019,Ayres2018,Fomin2018,Fomin2017,Serebrov2013MC} of UCN
losses taking into account the calculated incidence angles of each collision
for the particular trap geometry, provided the initial momentum distribution
of UCN is known and the applied simple physical model of UCN interaction
with trap walls is correct. The Monte-Carlo simulations are also actively
used to estimate the UCN losses in magnetic traps \cite{Callahan2019}. A
more serious problem is the surface roughness of material traps, which makes
impossible the exact calculation of UCN scattering angle and loss
probability during each collision.

The UCN losses on trap walls and the extrapolation interval can be reduced
by increasing the trap size and, hence, the volume-to-surface ratio in
material traps. However, even with a very large UCN trap with size 2~m in
the recent $\tau _\text{n}$ measurements \cite%
{Serebrov2018PhysRevC.97.055503} the extrapolation interval $\tau _\text{n}%
-\tau_\text{m}$ was only reduced to 20 seconds. A further size increase of
UCN traps is not only technically problematic but also not very useful,
because the main UCN losses already come from their collisions with the trap
bottom, and the rate of these collisions is determined by the Earth gravity
and by the UCN kinetic energy $E_\text{k} <V_0$, rather than by the trap
size.

A possible qualitative way to reduce the UCN absorption rate is to cover the
trap walls by liquid $^{4}$He, the only material that does not absorb
neutrons \cite%
{Golub1983,Bokun/1984,Alfimenkov/2009,Grigoriev2016Aug,GrigorievPRC2021,Grigoriev2021,GrigorievPRC2023}%
. However, covering the UCN trap walls by liquid helium has several
drawbacks. First, $^{4}$He creates a very small optical potential barrier $%
V_{0}^{\text{He}}=18.5\,\text{neV}$ %\ll V_{0}^{\text{F}} \approx 106$~neV 
for neutrons, which is $5.7$ times smaller than the barrier height $V_{0}^{%
\text{F}}\approx $ 106~neV of Fomblin oil. Hence, the UCN phase volume and
their density in such a trap is reduced by the factor $(V_{0}^{\text{F}%
}/V_{0}^{\text{He}})^{3/2}\approx 13.7$ as compared to the Fomblin coating,
which enhances the statistical errors. The UCN production technology
develops \cite{PhysRevC.99.025503,ZimmerPhysRevC.93.035503,Abele2023}, and
this reduction of neutron density may become less important than the
advantage of decreasing the UCN absorption rate. The second problem with the
liquid $^{4}$He coating of UCN trap walls is that a very low temperature $%
T<0.5$~K is required. At higher $T$ the concentration of $^{4}$He vapor is
rather high, leading to the inelastic UCN scattering with large energy
transfer $\sim k_\text{B}T\gg V_{0}^{\text{He}}$ from the vapor atoms to
neutrons. The third problem with liquid $^{4}$He is another source of
inelastic UCN scattering -- the thermally activated quanta of surface waves,
called ripplons. They lead to a linear temperature dependence of scattering
rate \cite{Grigoriev2016Aug}, surviving even at ultra-low temperature. The
strength of neutron-ripplon interaction is rather small \cite%
{Grigoriev2016Aug}, which makes feasible the UCN storage in He-covered
traps, and the linear temperature dependence of UCN losses due to their
scattering by ripplons is very convenient for taking into account this
systematic error. However, the UCN scattering by ripplons strongly limits
the possible advantage of using liquid helium in the UCN storage. Hence,
below we consider more traditional and currently used UCN traps, where the
wall material absorbs neutrons.

The ''neutron lifetime puzzle'', i.e. the difference between $\tau _\text{n}$%
-measurements using cold neutron beam and material UCN traps, is generally
attributed to the errors in beam experiments \cite%
{Serebrov2021PhysRevD.103.074010}. However, the 4-second discrepancy between
the results of latest magnetic-trap and material-bottle $\tau _\text{n}$
measurements remains puzzling. Notably, the former also very precise
measurements of UCN lifetime using a material trap gave the value \cite%
{Serebrov2008PhysRevC.78.035505} $\tau _\text{n}$ = 878.5 $\pm $ 0.7 (stat) $%
\pm $ 0.3 (sys) s, which is 3 seconds smaller than the result of Ref. \cite%
{Serebrov2018PhysRevC.97.055503} and much closer to the magnetic-trap $\tau _%
\text{n}$ measurements. The difference between Refs. \cite%
{Serebrov2018PhysRevC.97.055503} and \cite{Serebrov2008PhysRevC.78.035505}
is not only a larger size but also the different shape of UCN trap used in
the experiment \cite{Serebrov2018PhysRevC.97.055503}. 
As we show below, the
size extrapolation used to estimate the UCN loss rate $\tau _\text{loss}%
^{-1} $ due to the absorption by trap walls is rather sensitive to the trap
shape, which may give a 3-second difference in the measured $\tau _%
\text{n} $. In this paper we reanalyze the estimate method of UCN loss rate $%
\tau _\text{loss}^{-1}$ from the absorption by trap walls and consider
several important assumptions which may lead to errors in the extracted
value of neutron lifetime.

In Sec. \ref{SecII} we summarize the standard estimate procedures of UCN
loss rate in material traps and emphasize their approximations. In Sec. \ref%
{SecCalulations} we analytically calculate the UCN absorption rate for two
simple trap shapes, assuming the uniform and isotropic velocity distribution of UCN at trap bottom, 
and show the difference between various estimate methods
of UCN loss-rate calculations. In Sec. \ref{SecMonteCarlo} we performed 
 Monte-Carlo simulations of UCN distribution and absorption by trap walls.
In Sec. \ref{SecDiscussion} we discuss the results and their
consequences on the size scaling and on the accuracy of estimates of neutron
loss rate and lifetime.

\section{Standard estimate procedure of neutron losses in UCN traps and its
discussion}

\label{SecII}

\subsection{Basic formulas}

The methods of estimating $\tau_\text{loss}^{-1}$ start from the well-known
formula \cite{Golub/1991,Ignatovich/1990,Ignatovich1996} for the absorption
probability of neutron by the wall during each scattering 
\begin{equation}
\mu \left( v_{\perp }\right) = \frac{2\eta \ v_{\perp}/v_{\lim} }{\sqrt{%
1-(v_{\perp}/v_{\lim})^{2}}},  \label{mu}
\end{equation}
where $\eta$ is the loss coefficient depending on the wall material, $%
v_{\perp }$ is the normal-to-wall UCN velocity, and to the $v_{\lim }$ is
the limiting velocity, corresponding to the UCN kinetic energy $E_\text{k}$
equal to the potential barrier $V_{0} = m_\text{n} v_{\lim }^{2}/2$, created
by trap material. Then one usually assumes a uniform distribution of the
incidence angle of UCN on the walls. The integration of Eq. (\ref{mu}) over
the solid incidence angle gives another well-known formula \cite%
{Golub/1991,Ignatovich/1990,Ignatovich1996} for the averaged absorption
probability%
\begin{equation}
\bar{\mu} \left(v_*\right) =\frac{2\eta }{v_{* }^{2}}\left( \arcsin v_{*
}-v_{* }\sqrt{1-v_{* }^{2}}\right) \approx \left\{ 
\begin{array}{c}
\pi \eta ,~v_{* }\rightarrow 1 \\ 
4\eta v_{* }/3,~v_{* }\ll 1%
\end{array}
\right. ,  \label{mu(v)}
\end{equation}%
where the normalized velocity $v_{* } \equiv v/v_{\lim }=\sqrt{E_\text{k}%
/V_{0}}$. We rewrite Eq. (\ref{mu(v)}) as 
\begin{equation}
\bar{\mu} \left( E_*\right) =2\eta \ f_1 (E_{*}) ,  \label{mu(E)}
\end{equation}%
introducing the normalized UCN energy $E_{* }\equiv E_\text{k}/V_{0}=v_{*
}^{2}$ and the dimensionless function 
\begin{equation}
f_1 \left( x \right) = \frac{1}{x} \left( \arcsin\sqrt{x} - \sqrt{x}\sqrt{1-x%
} \right) ,  \label{f_1}
\end{equation}%
which describes the effective collision rate of UCN with trap walls.

\subsection{Gravity effects and size extrapolation}

The averaged absorption probability $\bar{\mu}$ depends on UCN kinetic
energy $E_\text{k}$, which due to the Earth gravity depends on the height $h$
above the trap bottom as $E_\text{k}(h) =E-m_\text{n} g h = E-h^{\prime}$,
where $E$ is total UCN energy, equal to the kinetic neutron energy at the
trap bottom $h=0$, %$m_{n}=$ $1.675\cdot 10^{-24}$ g is the neutron mass, 
$g$ is the free-fall acceleration, and $h^{\prime }\equiv m_\text{n}gh$. The
corresponding height dependence of neutron velocity $v_{* } (h) =\sqrt{(E-m_%
\text{n}gh) /V_{0}}$ is, usually, taken into account by the integration of
the averaged scattering rate $\bar{\mu}\left( E_\text{k} \right) $ over the
trap surface as \cite%
{Serebrov2008PhysRevC.78.035505,Serebrov2018PhysRevC.97.055503} 
\begin{multline}
\tau_{\text{(g)}}^{-1} \left( E\right) = \frac{\int_{0}^{h_{\max } (E)}
dS(h) \ \bar{\mu} ( E-h^{\prime }) v( E-h^{\prime }) \rho ( E,h^{\prime }) 
} {4\int_{0}^{h_{\max }(E) } dV(h) \ \rho (E,h^{\prime }) } \\
\equiv \eta \ \gamma (E) ,  \label{tau_g}
\end{multline}
where according to Ref. \cite{Serebrov2018PhysRevC.97.055503} the UCN number
density 
\begin{equation}
\rho \left( E,h\right) \propto \sqrt{\left( E-h^{\prime }\right) /E}
\label{rho}
\end{equation}%
gives the energy and height distribution of UCN in the trap, 
\begin{equation}
h_{\max }( E) \equiv E/m_\text{n}g = v^{2}/2g \equiv h_{\lim}E_{*}
\label{hmax}
\end{equation}
is the maximal height of neutrons with energy $E$, and $\gamma (E) $ is the
effective collision frequency of UCN with the walls. This effective UCN
collision frequency also enters the size extrapolation formula that gives
the neutron $\upbeta$-decay time $\tau _\text{n}$ from two measured
lifetimes $\tau _{1}$ and $\tau _{2}$ in two UCN traps of different size 
\cite{Serebrov2018PhysRevC.97.055503}:%
\begin{equation}
\tau _\text{n}^{-1}= \left. \tau _{1}^{-1}-\left( \tau _{2}^{-1}-\tau
_{1}^{-1}\right) \middle/ \left( \frac{\gamma _{2}(E) }{ \gamma_{1}(E) } -1
\right) \right. .  \label{geo_extra}
\end{equation}

For the traps of height $h\ll h_{\lim}$ one can simplify further, ignoring
the gravity effects and assuming the isotropic velocity distribution. The
ratio of effective collision frequencies is then taken as the ratio of
surface-to-volume ratios of two different traps \cite{Ignatovich/1990}: 
\begin{equation}
\frac{\gamma _{2}(E) }{\gamma _{1}(E) }= \left. \frac{S_{2}}{V_{2}} \middle/ 
\frac{S_{1}}{V_{1}}\right..  \label{g}
\end{equation}
According to Eq. (\ref{g}), one obtains the following simple formula for the
UCN loss rate 
\begin{equation}
\tau_{\text{(s)}}^{-1} \left(E\right) =\frac{\bar{\mu}\left( E\right)
v\left( E\right) S}{4V}.  \label{tau_s}
\end{equation}%
Below we compare these two results, given by Eqs. (\ref{tau_g}) and (\ref%
{tau_s}), with another estimations for several simple trap shapes.

\subsection{Problems with standard size extrapolation and with estimates of
UCN losses}

Evidently, Eq. (\ref{g}) is oversimplified and contradicts Eq. (\ref{tau_g}%
), because the two UCN traps are, usually, not geometrically similar. Even
if they are similar in shape and differ only in size, the gravity effects
violate the simple formula (\ref{g}). Therefore, in Refs. \cite%
{Serebrov2008PhysRevC.78.035505,Serebrov2018PhysRevC.97.055503} Eq. (\ref%
{tau_g}) was used to estimate $\gamma(E) $. However, the energy dependence
of $\tau_\text{(g)}^{-1}(E) $ and $\gamma(E)$ in Eq. (\ref{tau_g}) makes the
size extrapolation to be energy dependent too. This problem can be partially
solved by the integration of the result for $\tau_\text{(g)}^{-1} (E) $ over
energy with the actual energy distribution $n(E) $ of UCN. Then the final $%
\tau_\text{n}$ result (\ref{geo_extra}) depends on the distribution function 
$n(E) $. This function is unknown. In Ref. \cite%
{Serebrov2018PhysRevC.97.055503} it was initially assumed Maxwellian and
then corrected by fitting the energy spectrum of UCN reaching the detector,
which can be measured. However, even this fitting procedure does not ensure
that the distribution function $n(E)$ is found precisely.

Another important problem is that Eqs. (\ref{mu(E)}) and (\ref{tau_g}) are
also approximate, because the gravity changes not only the neutron energy, as described by Eq. (\ref{tau_g}), 
but also the velocity distribution, because it affects only the vertical $z$%
-component of UCN velocity. %This has two main effects on the estimate procedure of UCN losses. First
Hence, the assumption of uniform angular
distribution of UCN velocities, implied in Eqs (\ref{mu(E)}) and (\ref{tau_g}%
), even if holds at the trap bottom, may violate on the side trap walls at $%
h\neq 0$. If this anisotropy of velocity distribution appears, it can be
compensated by a proper choice of the function $\bar{\mu}^{*}(y) $, which
then depends on the trap size and geometry. By the Monte-Carlo
simulations it was shown \cite{Serebrov2018PhysRevC.97.055503} that while
the result of energy extrapolation is very sensitive to the function $\bar{%
\mu}(y) $ and differs by 40 seconds for different $\bar{\mu}(y)$, the result
of subsequent size extrapolation is much less sensitive to the function $%
\bar{\mu}(y)$ and varies withing the interval of only 4 seconds. 
%However, these MC similations themselves were based on Eqs. (\ref{mu(E)}) and (\ref{tau_g}) and do not take
%into account the possible dependence of the optimal $\bar{\mu}^{*}(y)$ on the trap geometry, which spoils the size extrapolation.
Another important issue to be considered is the proper dependence of the collision rate on UCN velocity. 
Even if the UCN velocity distribution is isotropic, for each neutron the dependence
of collision rate on its vertical velocity strongly differs from that
for horizontal, as follows from Eq. (\ref{N}) below. Then the effective
collision rate is not given by Eqs. (\ref{mu(v)})-(\ref{f_1}), which affects
the accuracy of UCN loss estimates.  Of course, the diffuse UCN scattering by trap walls 
complicates the above arguments. But can it restore the validity of Eq. (\ref{tau_g})?

There is also a more fundamental question about the validity of Eq. (\ref{mu})
for the neutron absorption rate during the collisions, especially for the
walls with imperfections as pores or rough unflat surface. This problem can
be partially solved covering the trap wall by a liquid film, as $^{4}$He in
the proposal of Refs. \cite%
{Golub1983,Bokun/1984,Alfimenkov/2009,Grigoriev2016Aug,GrigorievPRC2021,Grigoriev2021,GrigorievPRC2023}%
. However, the typical experiments are performed at higher temperature and
with solid trap walls, e.g. the Fomblin grease \cite%
{Serebrov2008PhysRevC.78.035505,ArzumanovPhysLettB2015,Serebrov2017,Serebrov2018PhysRevC.97.055503,
Review2019Pattie,MAMBOPhysRevLett.63.593,PICHLMAIER2010}%
. In this paper we assume Eq. (\ref{mu}) to be valid, and for several simple
UCN trap shapes we analyze the effect of the above mentioned anisotropy of
UCN velocity distribution and of effective collision frequency on the UCN
loss rate and on the accuracy of $\tau _{\text{n}}$ extraction.

%\section{Analytical formulas for specular UCN reflections in a rectangular trap}
\section{Analytical formulas for UCN losses in rectangular and cylindrical traps}

\label{SecCalulations}

%\subsection{The method of calculations}

\subsection{General formulas}

For a rectangular UCN trap of size $L_{x}\times L_{y}\times L_{z}$ the
neutron motion along three main axes separate, because any elastic
scattering by the trap wall perpendicular to axis $i$ only changes the sign
of neutron velocity $v_{i}$ along this axis. The absorption probability
during each collision is given by Eq. (\ref{mu}) and also depends only on
the same velocity component $v_{i}=v_{\perp }$:%
\begin{equation}
\mu _{i}\left( v_{i}\right) = \frac{2\eta v_{i}/v_{\lim }}{\sqrt{%
1-v_{i}^{2}/v_{\lim }^{2}}}.  \label{mu_i}
\end{equation}%
The number of collisions with the walls during a long time $t\gg L_{i}/v_{i}$
for each neutron can be easily estimated: 
\begin{equation}
\mathcal{N}_{x}=\frac{t v_{x}}{L_{x}}, ~\mathcal{N}_{y}=\frac{t v_{y}}{L_{y}}%
, ~\mathcal{N}_{z}= \frac{t g}{2 v_{z}},  \label{N}
\end{equation}%
where the vertical velocity component $v_{z}$ is taken at the trap bottom $%
z=0$. From Eq. (\ref{N}) we already see a strong difference between the
dependence of UCN collision rate on their vertical and horizontal
velocities. While the collision rate with side walls is proportional to the
horizontal UCN velocity and is not affected free fall acceleration, the
collision rate with trap bottom is inversely proportional to the vertical
UCN velocity, because there is no upper trap wall, and the neutrons fall to
the bottom only because of gravity. Hence, there is a big difference between
the collisions with trap bottom and side walls, which affects the usual
procedure of size extrapolation of UCN loss rate to an infinitely large UCN
trap.

The probability for a neutron to remain in the trap after time $t$ is given
by the product 
\begin{equation}
P_{\text{n}}\left( t,\bm{v}\right) =P(t,\bm{v})e^{-t/\tau _{\text{n}%
}},  \label{Pn}
\end{equation}%
where the probability for a neutron to be not absorbed by the trap walls is 
\begin{equation}
P\left( t,\bm{v}\right) =\Pi _{i=x,y,z}\left[ 1-\mu _{i}(v_{i})\right] ^{%
\mathcal{N}_{i}}.  \label{P}
\end{equation}%
Since $\mu _{i}(v_{i})\sim \eta \ll 1$ and $\mathcal{N}_{i}\sim \eta
^{-1}\gg 1$, using $e=\lim_{n\rightarrow \infty }(1+1/n)^{n}$ one may
simplify Eq. (\ref{P}) to 
\begin{equation}
P\left( t,\bm{v}\right) \approx \exp \left( -\sum_{i}\mathcal{N}_{i}\mu
_{i}\left( v_{i}\right) \right) =\exp \left( -\frac{t}{\tilde{\tau}_{\text{%
(e)}}\ (\bm{v})}\right) ,  \label{Pa}
\end{equation}%
where the exact absorption rate 
\begin{equation}
\tilde{\tau}_{\text{(e)}}^{-1}\left( \bm{v}\right) =\sum_{i}\frac{\mu
_{i}(v_{i})\mathcal{N}_{i}}{t}.  \label{tau}
\end{equation}

Eqs. (\ref{tau}) and (\ref{tau(E)}) should be compared with the generally
used formulas for $\tau _{\text{(g)}}^{-1}(E)$, given by Eqs. (\ref{mu(E)}%
)--(\ref{rho}), and with the oversimplified formula for $\tau _{\text{(s)}%
}^{-1}(E)$, given by Eq. (\ref{tau_s}). The results differ because (i) the
probability (\ref{Pa}) depends not only on the velocity absolute value as in
Eq. (\ref{mu(E)}), but also on its direction, which is not isotropic at
finite height $h$, and (ii) according to Eq. (\ref{N}) the collision
frequency with the trap bottom $\mathcal{N}_{z}\propto v_{z}^{-1}$ is
inversely proportional to the normal velocity, contrary to the collision
frequency with the side walls $\mathcal{N}_{x}\propto v_{x}$ and $\mathcal{N}%
_{y}\propto v_{y}$. Below we analytically calculate the UCN absorption rate
by the walls of rectangular and cylindrical traps by these three methods and
compare the results obtained.

\subsection{Isotropic UCN velocity distribution}

If the physical model described by Eq. (\ref{mu_i}) is correct, Eqs. (\ref%
{tau}) and (\ref{N}) are exact for calculating the UCN absorption rate of
each neutron in a rectangular trap provided it velocity $\bm{v}$ at trap
bottom is known. However, to use these formulas one also needs to know the
UCN velocity distribution function $f_{0}\left( \bm{v}\right) $ at trap
bottom. Then the total UCN absorption rate%
\begin{equation}
\tau _{\text{(e)}}^{-1}=\int d^{3}\bm{v} \, \tilde{\tau}_{\text{(e)}}^{-1}(\bm{v}%
)f_{0} (\bm{v}) .
\label{tauE1}
\end{equation}%
The energy dependence of the UCN absorption rate 
\begin{equation}
\tau _{\text{(e)}}^{-1}\left( E\right) =\int d^{3}\bm{v}\,\delta \left(
E-\frac{m\bm{v}^{2}}{2}\right) \tilde{\tau}_{\text{(e)}}^{-1}(\bm{v}%
)f_{0} (\bm{v}) .  \label{tau(E)G}
\end{equation}%
Eqs. (\ref{tauE1}) and (\ref{tau(E)G}) do not contain the kinematic factor $\cos \theta $ as
in the calculation of collision rate (\ref{mu(v)}) because this factor is already contained in the
absorption rate $\tilde{\tau} _{\text{(e)}}^{-1}$.

\begin{figure}[tb]
    \centering
	\includegraphics[width=\linewidth]{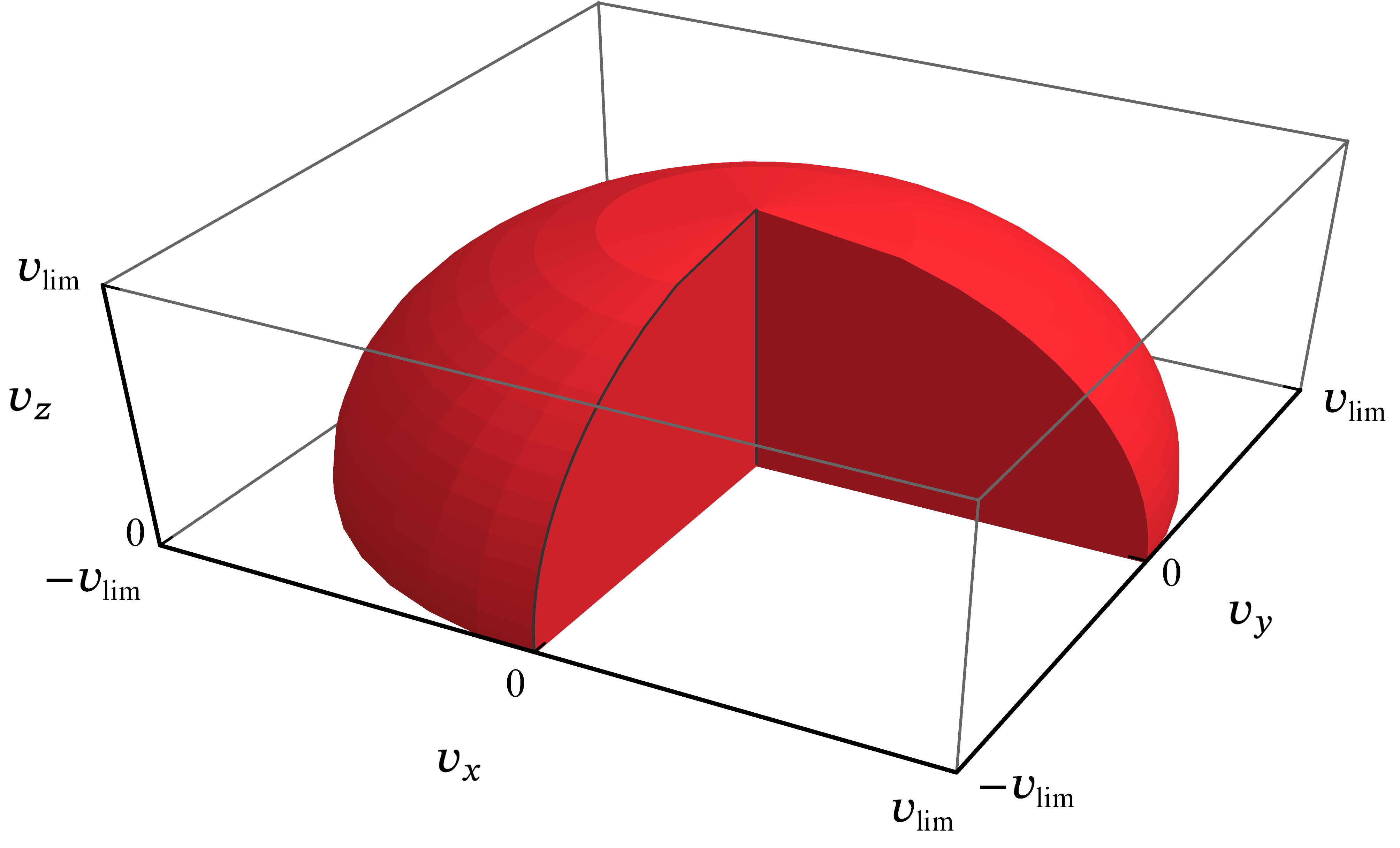}
    \caption{ Color plot of initial isotropic UCN velocity distributions. All velocities in hemisphere
	of radius $v_\text{lim}$ are equally probable.
    }
\label{FigIsotropic}
\end{figure}

One often assumes an isotropic and uniform \emph{initial} velocity
distribution 
\begin{equation}
f_{0}\left( v_{x},v_{y},v_{z}\right) =f_{0}= \const  \label{f0}
\end{equation}%
of neutrons at the trap bottom with the condition $\left\vert \bm{v}%
\right\vert \leq v_{\max }$, i.e. the uniform UCN distribution in the
velocity space limited by the sphere of radius $v_{\max }$ and volume $%
V_{v}=4\pi v_{\max }^{3}/3$. This isotropic UCN velocity distribution is 
illustrated in Fig. \ref{FigIsotropic}. Usually, in UCN-$\tau$ experiments 
 \cite{Serebrov2008PhysRevC.78.035505,Serebrov2018PhysRevC.97.055503} $v_{\max }<0.9\,v_{\lim }$. 
The normalization condition states that the total neutron number 
\begin{equation}
N_{0}=S_{b}\int_{\left\vert \bm{v}\right\vert \leq v_{\max }}d^{3}%
\bm{v\,}f_{0}\left( v_{x},v_{y},v_{z}\right) =S_{b}f_{0}\frac{4\pi }{3}%
v_{\max }^{3},  \label{NNorm}
\end{equation}%
where $S_{b}$ is the area of trap bottom. 
%The number of UCN with energy $E\leq E_{\ast }V_{0}$ is $N( E_{\ast }) =N_{0}E_{\ast }^{3/2}$,
%and the reduced UCN density of states $n (E_{\ast }) =dN (E_{\ast }) /dE_{\ast } =3N_{0}\sqrt{E_{\ast }}/2$. 
For isotropic $f_{0} (\bm{v})$ the energy dependence of the UCN absorption rate is given by the integration 
over the solid angle: 
\begin{equation}
\tau _{\text{(e)}}^{-1}\left( E\right) =\int \frac{d\varOmega}{4\pi }\tilde{%
\tau}_{\text{(e)}}^{-1}(\bm{v}).  \label{tau(E)}
\end{equation}
%Eq. (\ref{tau(E)}) does not contain the kinematic factor $\cos \theta $ as in the calculation of collision rate because it is already contained in the absorption rate $\tau _{\text{(e)}}^{-1}$. 
The isotropic velocity distribution (\ref{f0}) at trap bottom does not mean
that at any height the UCN velocity distribution remains isotropic, because
with height change the UCN vertical velocity also changes due to gravity. %
\iffalse At $z=0$ the neutrons of total number $N_{0}$ uniformly fill the
semisphere of radius $v_{\max }$. The number $N(h)$ of neutrons reaching the
height $h<h_{\max }$ is proportional to the volume $V_{s}=\pi
v_{h}^{2}\left( v_{\max }-v_{h}/3\right) $ of spherical segment of height $%
v_{h}=v_{h}\left( h\right) =v_{\max }-\sqrt{2gh}$: 
\begin{eqnarray}
\frac{N(h)}{N_{0}} &=&\frac{V_{s}}{V_{v}/2}=\frac{\pi \left( v_{\max }-\sqrt{%
2gh}\right) ^{2}\left( 2v_{\max }+\sqrt{2gh}\right) }{2\pi v_{\max }^{3}} 
\notag \\
&=&\left( 1-\sqrt{h/h_{\max }}\right) ^{2}\left( 1+\sqrt{h/h_{\max }}%
/2\right) .
\end{eqnarray}%
\fi

One should distinguish between the total UCN \emph{number} distribution
function $f_{0}( \bm{v}) $ and the \emph{density}
distribution $n( h,\bm{v}) $ on the height $h$, which gives
the UCN density in the 6-dimensional phase space. Below we show that the
isotropic velocity distribution $f_{0}( \bm{v}) $ results to
anisotropic $n( h,\bm{v}) $ and vice versa even at the trap
bottom $h=0$. If $n( h,\bm{v}) $ is known, the UCN velocity
distribution at trap bottom is given by 
\begin{equation}
f_{0}\left( v_{x},v_{y},v_{z}\right) =\int_{0}^{h_{\max }}dh\,n\left(
h,v_{x},v_{y},\sqrt{v_{z}^{2}-2gh}\right) .
\end{equation}%
The Liouville's theorem asserts that the phase-space distribution function $%
\propto n ( h,\bm{v}) $ is constant along the trajectories of
the system, but $n ( h,\bm{v}) \neq \const$. Moreover, a diffuse elastic UCN 
scattering violates the Liouville's theorem. The diffuse
scattering obeying the Lambert's cosine law keeps the $n ( h,\bm{v})$ 
rather than $f_{0} (\bm{v}) $ isotropic, as we show in Sec. \ref{SecMonteCarlo}.

\subsection{Limiting case of a wide trap with uniform velocity distribution
at trap bottom}

First, we consider a limiting case of very wide UCN trap with $%
L_{x},L_{y}\gg h_{\max }$, where the neutron scattering happens only with
the trap bottom at $h=0$. We also assume isotropic and uniform %\emph{initial}
velocity distribution at trap bottom, given by Eqs. (\ref{f0}) and (\ref{NNorm}).

\subsubsection{Distribution function at arbitrary height}
\label{SubSecDistr}

The distribution function $f( h,\bm{v}) $ of neutrons on
height $h$ can be obtained from the initial uniform distribution function $%
f( h=0,\bm{v}) =f_{0}( \bm{v}) =f_{0}(v_{x},v_{y},v_{z}) $ at $h=0$ by reducing 
the velocity $z$-component as 
\begin{equation}
v_{z}\rightarrow v_{zh}\equiv v_{z}\left( h\right) =\sqrt{v_{z}^{2}-2gh}%
,~v_{z}=\sqrt{v_{zh}^{2}+2gh}  \label{vzh}
\end{equation}%
which gives the derivatives 
\begin{equation}
\frac{dv_{zh}}{dv_{z}}=\frac{v_{z}}{\sqrt{v_{z}^{2}-2gh}},\frac{dv_{z}%
}{dv_{zh}}=\frac{v_{zh}}{\sqrt{v_{zh}^{2}+2gh}},~~
\end{equation}%
and the particle velocity distribution at height $h$ 
\begin{eqnarray}
f\left( h,\bm{v}\right) \equiv f\left( v_{x},v_{y},v_{zh}\right)
=f_{0}\left( v_{x},v_{y},v_{z}\right) \frac{dv_{z}}{dv_{zh}} \\
=f_{0}\left( v_{x},v_{y},\sqrt{v_{zh}^{2}+2gh}\right) \frac{v_{zh}}{\sqrt{%
v_{zh}^{2}+2gh}}=f_{0}\frac{v_{zh}}{v_{z}}.  \label{fvzh}
\end{eqnarray}%
However, this distribution function is not proportional to neutron density
distribution $n( \bm{v}) $ on the height $h$ or, more
precisely, in the height interval $\left( h,h+\Delta h\right) $ at $\Delta
h\rightarrow 0$, because the latter is also proportional to the ratio of the
time 
\begin{equation}
\Delta t\left( \bm{v}\right) =\Delta t\left( v_{z},h\right) =\Delta
h/\left\vert v_{zh}\right\vert  \label{dt}
\end{equation}%
that each neutron spends on the height $h$ to the total time $t_{tr}=v_{z}/g$
on its trajectory. \iffalse More rigorously, $\Delta t\left( v_{z}\right)
=t_{2}-t_{1}$, where $-gt_{1}^{2}/2+v_{z}t_{1}=h$ and $%
-gt_{2}^{2}/2+v_{z}t_{2}=h+\Delta h$, i.e. the times $t_{1},t_{2}$ are when
the neutron enters and leaves the thin slab $h\in \left( h,h+\Delta h\right) 
$. This gives $gt_{1}^{2}/2-gt_{2}^{2}/2+v_{z}t_{2}-v_{z}t_{1}=\Delta h$, or 
$\Delta t\left( v_{z}\right) =\Delta h/\left[ v_{z}-g\left(
t_{2}+t_{1}\right) /2\right] =\Delta h/\left\vert v_{zh}\right\vert $.\fi %
Thus we obtain the particle density distribution at height $h$%
\begin{eqnarray}
n\left( h,\bm{v}\right) &=&\frac{\Delta N}{S_{b}\Delta h}=\frac{f\left(
h,\bm{v}\right) }{v_{zh}\left( v_{z}/g\right) }=\frac{f_{0}}{v_{z}^{2}/g}
\label{nhv} \\
&=&f_{0}\left( v_{x},v_{y},\sqrt{v_{zh}^{2}+2gh}\right) \frac{g}{%
v_{zh}^{2}+2gh}.  \label{nhv1}
\end{eqnarray}%
If $f_{0}( \bm{v}) =\const$, the particle density $n( h,\bm{v}) \neq \const$. 
At a large height $h\rightarrow h_{\max}=v_{\max }^{2}/2g$, where 
$2gh\gg v_{zh}^{2}$ and $v_{zh}^{2}+2gh\approx
2gh $, Eq. (\ref{nhv1}) gives an almost uniform and isotropic velocity
distribution $n( h,\bm{v}) $. However, at small height $h\ll
h_{\max }$ we obtain an \emph{anisotropic} velocity distribution, peaked at
small $\vert v_{zh} \vert $. At $h=0$ we have a singularity $%
n( h=0,\bm{v}) \propto 1/v_{zh}^{2}=1/v_{z}^{2}$. In the UCN
absorption rate by trap bottom this singularity of UCN density distribution
at $h=0$ cancels, because the collision rate $\mathcal{N}_{z}(
v_{z}) \propto $ $n( h=0,\bm{v}) v_{z}\approx
\const/v_{z}$, and another factor $v_{z}$ comes from the absorption
probability $\mu $ given by Eq. (\ref{mu}), as we see below.

\subsubsection{Absorption rate calculated from using the particle flow on a wall}

The absorption rate $dN/dt$ by trap walls is, usually, calculated by
integrating the particle flow on a wall, equal to the collision rate 
$\mathcal{N}_{c}( \bm{v}) =n( h,\bm{v}) v_{\perp}$, weighted by the absorption 
probability $\mu (v_{\perp})$ over all UCN velocities. Here $n( h,\bm{v})$ 
is the velocity distribution of UCN density on the height $h$, and $v_{\perp}$ 
is the velocity component perpendicular to the wall. 
This approach results \cite{Ignatovich/1990,Golub/1991} to Eq. (\ref{mu(v)}) 
and to Eq. (\ref{tau_g}) for isotropic $n( h,\bm{v})$. 
Substituting Eqs. (\ref{mu}) and anisotropic distribution (\ref%
{nhv}) we obtain the absorption rate $dN( E_{\ast }) /dt$ by trap
bottom of UCN with energy $E\leq E_{\ast }V_{0}$ 
\begin{eqnarray}
\frac{dN\left( E_{\ast }\right) }{dt} &=&\int_{v_{z}<0}d^{3}\bm{v} \,
n( h=0,\bm{v}) v_{z}\mu (v_{z})  \label{dNt0} \\
&=&\int_{v_{z}<0}d^{3}\bm{v}\, \frac{2f_{0}g\eta /v_{\lim }}{\sqrt{%
1-v_{z}^{2}/v_{\lim }^{2}}}.  \notag
\end{eqnarray}%
At each $v_{z}$ the volume in velocity space filled by UCN is 
$dV=\pi (v^{2}-v_{z}^{2}) dv_{z}$, which gives 
\begin{eqnarray}
\frac{dN\left( E_{\ast }\right) }{dt} &=&2\,\pi \eta \ f_{0}gv_{\lim
}^{2}\int_{0}^{v_{\ast }}dv_{z\ast }\frac{v_{\ast }^{2}-v_{z\ast }^{2}}{%
\sqrt{1-v_{z\ast }^{2}}}  \notag \\
&=&\frac{3\ N_{0}\eta g\,/v_{\lim }}{2(v_{\max }/v_{\lim })^{3}}\int_{0}^{%
\sqrt{E_{\ast }}}dx\frac{E_{\ast }-x{}^{2}}{\sqrt{1-x^{2}}},
\end{eqnarray}%
where $v_{z\ast }=$ $v_{z}/v_{\lim }$ and in the last line we used the
normalization of $\ f_{0}=N_{0}/( 4\pi v_{\max }^{3}/3) $. Taking
the integral we obtain the UCN loss rate of neutrons with energy less than $%
E_{\ast }V_{0}$: 
\begin{equation}
\dot{N}\left( E_{\ast }\right) =\frac{dN\left( E_{\ast }\right) }{dt}=\frac{%
3\ N_{0}\eta g\,/v_{\lim }}{2\left( E_{\max }/V_{0}\right) ^{3/2}}f_a(E_{\ast }),  
\label{NE}
\end{equation}%
where 
\begin{equation}
f_a\left( E_{\ast }\right) =\sqrt{E_{\ast }}\frac{\sqrt{1-E_{\ast }}}{2}%
+\left( \sqrt{E_{\ast }}-\frac{1}{2}\right) \arcsin  \sqrt{E_{\ast }}.  \label{fE}
\end{equation}%
One sees that at $E_{\ast }\rightarrow 0$ this loss rate 
$\dot{N}(E_{\ast }) \propto E_{\ast }\rightarrow 0$: 
\begin{equation*}
f_a\left( E_{\ast }\right) \approx E_{\ast }-E_{\ast }^{3/2}/3+E_{\ast
}^{2}/6+\dots,~E_{\ast }\rightarrow 0.
\end{equation*}%
To calculate the total UCN loss rate one only needs to substitute $E_{\ast
}=E_{\max }/V_{0}$ to Eq. (\ref{NE}).

\subsubsection{Alternative way of calculating the UCN absorption rate}

We can also calculate the UCN absorption rate using Eq. (\ref{tau}). For the
scattering by trap bottom we have $i=z$ and%
\begin{equation}
\tilde{\tau}_{\text{(e)}}^{-1}\left( \bm{v}\right) =\sum_{i}\frac{\mu
_{i}(v_{i})\mathcal{N}_{i}}{t}=\frac{\eta g/v_{\lim }}{\sqrt{%
1-v_{z}^{2}/v_{\lim }^{2}}}.  \label{TauEs}
\end{equation}%
The total loss rate of UCN 
\begin{equation}
\frac{dN}{dt}=-N_{0}\left\langle \frac{dP\left( t,\bm{v}\right) }{dt}%
\right\rangle =N_{0}\left\langle \tilde{\tau}_{\text{(e)}}^{-1}\left( 
\bm{v}\right) P\left( t,\bm{v}\right) \right\rangle ,  \label{dNt}
\end{equation}%
where the triangular brackets mean the averaging over all UCN. At $t=0$ $%
P\left( t,\bm{v}\right) =1,$ and combining Eqs. (\ref{TauEs}) and (\ref%
{dNt}) we have at $t=0$ 
\begin{equation}
\frac{dN}{dt}=\frac{N_{0}}{4\pi v_{\max }^{3}/3}\int d^{3}\bm{v}\frac{%
\eta g/v_{\lim }}{\sqrt{1-v_{z}^{2}/v_{\lim }^{2}}}.
\end{equation}%
This coincides with Eq. (\ref{dNt0}) and gives the same result. Hence, the above two methods of calculating the UCN absorption rate are equivalent provided the UCN distribution is taken the same.

\iffalse

The Liouville's theorem asserts that the phase-space distribution function $%
\propto n ( h,\bm{v}) $ is constant along the trajectories of
the system, but $n ( h,\bm{v}) \neq \const$. Moreover, a diffuse elastic UCN 
scattering violates the Liouville's theorem. The diffuse
scattering obeying the Lambert's cosine law keeps the $n ( h,\bm{v})$ 
rather than $f_{0} (\bm{v}) $ isotropic, as we show in Sec. \ref{SecMonteCarlo}.
Eq. (\ref{nhv1}) does not contradict the Liouville's theorem of constant
particle density $n( h,\bm{v}) $ in the phase space along the
trajectory, because it gives the initial UCN height-velocity distribution.
Due to the diffuse scattering by trap bottom obeying the Lambert cosine law,
the velocity distribution of UCN density at trap bottom $n( h=0,\bm{v}) $ 
becomes isotropic after some time even if it was initially anisotropic. 
Hence, one may expect that Eqs. (\ref{nhv}) and (\ref{f0}) do not hold if the diffuse scattering is included. 
\fi

\subsection{Comparison of the absorption rate by trap bottom calculated by
different methods for isotropic UCN number velocity distribution}

In this subsection we again consider a very wide UCN trap with $L_{x},L_{y}\gg h_{\max }$,
where the neutron scatter mainly by the trap bottom at $h=0$, and calculate the energy 
dependence of the absorption rate by three different methods. This may be useful if 
the energy distribution of UCN is known and differs from the one corresponding to 
uniform velocity distribution in Eq. (\ref{f0}), as in real experiments \cite{Serebrov2008PhysRevC.78.035505,Serebrov2017}.

Eqs. (\ref{mu(E)})--(\ref{rho}), which take the gravity into account but
assume (i) an isotropic UCN velocity distribution at any height and (ii)
contrary to Eq. (\ref{N}), the similar dependence of the effective collision
rate on the vertical and horizontal UCN velocity, give%
\begin{equation}
\tau_{z\text{(g)}}^{-1} \left(E\right) =\frac{m_\text{n} g \ \bar{\mu} (E) v 
}{ 4\int_{0}^{E}dh^{\prime }\sqrt{1-h^{\prime }/E}} =\frac{3}{2}
\tau_{z0}^{-1} \frac{f_1\left( E_{* }\right) }{\sqrt{E_{* }}}.
\label{tau_zg}
\end{equation}
Here we introduced the factor $\tau_{z0}^{-1}= g \eta / v_{\lim}$, which
describes the UCN absorption rate by trap bottom in order of magnitude, and
the function $f_1\left( E_{* }\right) $ is given by Eq. (\ref{f_1}).

When the gravity effects are neglected and an isotropic velocity
distribution is assumed, for the size extrapolation in $\tau_\text{n}$
measurements one often applies Eq. (\ref{tau_s}) for calculating the UCN
loss rate $\tau_{\text{(s)}}^{-1}(E) $, which just gives a geometrical
surface-to-volume ratio of the UCN trap. This oversimplified method
neglecting the gravity seems to be completely inapplicable for the
calculation of UCN loss rate due to the collisions with trap bottom, because
the corresponding collision rate $\mathcal{N}_{z}=g /2v_{z}$ in Eq. (\ref{N}%
) is determined by gravity. Nevertheless, if we take as the ''trap height'' $%
V/S$ in Eqs. (\ref{g}) and (\ref{tau_s}) the maximum UCN height in the
gravitational potential $h_{\max }(E)$, given by Eq. (\ref{hmax}), then we
obtain the following simple formula for the UCN absorption rate 
\begin{equation}
\tau_{z\text{(s)}}^{-1} \left(E\right) =\tau_{z0}^{-1} \frac{f_1( E_{*
}) }{\sqrt{ E_{* }}}.  \label{tau_zs}
\end{equation}%
It differs from $\tau_{z\text{(g)}}^{-1}(E) $ only by the factor $2/3$,
which came from the integral $\int_{0}^{1}dx\sqrt{1-x}=2/3$ in the
denominator of Eq. (\ref{tau_zg}). Physically, this difference comes because
Eq. (\ref{tau_g}) takes into account the dependence of UCN density on
height, given by Eq. (\ref{rho}), which is maximal at the trap bottom and, hence, gives a larger
absorption rate by the trap bottom, while Eq. (\ref{g}) assumes a uniform
UCN density.

\begin{figure}[tb]
\label{FigTauZ}\centering
\includegraphics{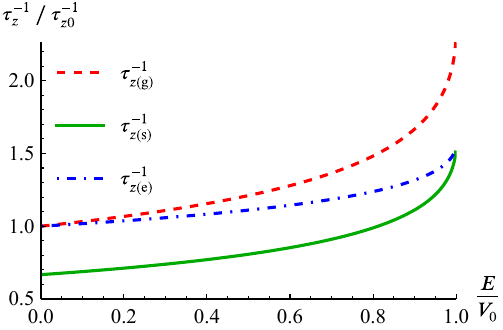}
\caption{The energy dependence of UCN loss rate $\protect\tau_{z}$ from the
absorption by trap bottom only, corresponding to a wide UCN trap. The
calculations and averaging over UCN incidence angle are done in three ways:
(i) the standard method, using Eqs. (\protect\ref{mu(E)})--(\protect\ref{rho}%
) and resulting to Eq. (\protect\ref{tau_zg}) (dashed red line), which
assumes that the gravity only changes the UCN density and velocity absolute
value (as a function of height) but not its angular distribution, (ii) the
oversimplified method, neglecting all gravity effects and giving Eq. (%
\protect\ref{tau_zs}) (solid green line), and (iii) the exact calculation
method for the isotropic UCN velocity distribution at trap bottom given by Eq. (\ref{f0}) 
and illustrated in Fig. \ref{FigIsotropic}, giving Eq. (\protect\ref{tau_za}) (dot-dashed blue
line). }
\end{figure}

The calculation using Eqs. (\ref{mu_i}), (\ref{N}), (\ref{tau}) and
(\ref{tau(E)}) for an isotropic velocity distribution (\ref{f0}) of UCN near the trap
bottom result to a different UCN absorption rate 
\begin{equation}
\tau_{z\text{(e)}}^{-1} \left(E\right) =\tau_{z0}^{-1} \frac{\arcsin \sqrt{%
E_{* }} }{\sqrt{E_{* }}}.  \label{tau_za}
\end{equation}%
Eqs. (\ref{tau_zg}) and (\ref{tau_za}) give the same value of the UCN
absorption rate at $E=0$ but differ considerably at $E>0$, which follows
already from their Taylor expansions at $E_* = E/V_0 \ll 1$: 
\begin{equation}
\frac{ \tau_{z\text{(g)}}^{-1} } { \tau_{z0}^{-1} } \simeq 1+\frac{3}{10}%
E_*, \quad \frac{ \tau_{z\text{(e)}}^{-1} } { \tau_{z0}^{-1} } \simeq 1+%
\frac{E_*}{6}.
\end{equation}
The oversimplified result in Eq. (\ref{tau_zs}) gives different both the UCN
absorption rate at $E=0$ and its asymptotics: 
\begin{equation}
\frac{ \tau_{z\text{(s)}}^{-1} } { \tau_{z0}^{-1} } \simeq \frac{2}{3}+\frac{%
E_*}{5}.
\end{equation}
The analytical results (\ref{tau_zg}), (\ref{tau_zs}) and (\ref{tau_za}) are
plotted in Fig. \ref{FigTauZ}. They hold for the UCN absorption rate by the
trap bottom not only for rectangular but for any straight cylindrical trap
with an arbitrary base shape.

\subsection{Absorption by rectangular side walls}

\label{SecRectangularSideWalls}

Now we consider the UCN absorption by side walls calculated using the above
three methods. The standard extrapolation procedure described by Eqs. (\ref%
{mu(E)})--(\ref{rho}) for the rectangular trap of dimensions $L_{x} \times
L_{y}$ gives the following UCN absorption rate by side walls 
\begin{multline}
\tau_{\text{r(g)}}^{-1} = \frac{L_{x}+L_{y}}{2L_{x}L_{y}} \frac{%
\int_{0}^{h_{\max}(E) } dh \ \bar{\mu} (E-h^{\prime}) v(E-h^{\prime})
\rho(E,h^{\prime})} {\int_{0}^{h_{\max}(E) } dh \ \rho (E,h^{\prime})} \\
= 3 \tau_{r0}^{-1} \frac{f_2(E_*)}{\sqrt{E_*}} ,  \label{tau_rg}
\end{multline}%
where we introduced the factor $\tau_{r0}^{-1} = \eta v_{\lim} (L_{x}+L_{y})
/ 2 L_{x}L_{y}$ giving the UCN absorption rate by side trap walls in order
of magnitude, and the dimensionless function % \begin{multline}
% 	\! f_2 (y) 
% 	= \! \int_{0}^{1} \!\! dx 
% 	\left( \arcsin \! \sqrt{y (1-x)}-\sqrt{y (1-x) }\sqrt{1-y (1-x) }\right) \\
% 	= \left( \frac{3}{4y}-\frac{1}{2}\right) \! \sqrt{y(1-y)}
% 	+ \left( \frac{3}{4y}-1\right) \! \left( \arctan \! \sqrt{y^{-1}-1}-
% 	\frac{\pi }{2}\right). \label{f_2}
% \end{multline}
\begin{multline}
\!\!\! f_2 (y) = \! \int_{0}^{1} \!\! dx \left( \arcsin \! \sqrt{y (1-x)}-\sqrt{%
y (1-x) }\sqrt{1-y (1-x) }\right) \\
= \left( \frac{3}{4y}-\frac{1}{2}\right) \sqrt{y(1-y)} - \left( \frac{3}{4y}%
-1\right) \arcsin \sqrt{y}.  \label{f_2}
\end{multline}

Applying Eq. (\ref{tau_s}), which neglects gravity, we get the following
oversimplified absorption rate by side walls 
\begin{equation}
\tau_{r\text{(s)}}^{-1}\left( E\right) =2 \tau_{r0}^{-1} \sqrt{E_{*}} f_1
(E_{*}).  \label{tau_rs}
\end{equation}%
%
%
%
%
%
%
%
%
%where we used Eq. (\ref{mu(E)}) for $\bar{\mu}\left( E\right) $. 

Combining Eqs. (\ref{mu_i}),(\ref{N}),(\ref{tau}),(\ref{tau(E)}%
) we get the absorption rate by side walls due to the UCN horizontal motion for the uniform velocity distribution (\ref{f0}) of UCN at trap bottom:
%\textcolor{red}{!}
%\begin{eqnarray}
%\bar{\tau}_{La}^{-1} &=&\int_{0}^{\pi }d\theta \frac{\sin \theta }{2}%
%\int_{0}^{2\pi }\frac{d\phi }{2\pi }\frac{L_{x}+L_{y}}{L_{x}L_{y}}\frac{%
%2\eta v_{\lim }v_{* }^{2}\left( \sin \theta \cos \phi \right) ^{2}}{\sqrt{%
%1-v_{* }^{2}\left( \sin \theta \cos \phi \right) ^{2}}}  \notag \\
%&=&\eta v\frac{L_{x}+L_{y}}{L_{x}L_{y}}f_1\left( E_{* }\right) =\frac{%
%L_{x}+L_{y}}{2L_{x}L_{y}}2\eta v_{\lim }\sqrt{E_{* }}f_1\left( E_{*
%}\right),  %\label{tau_ra}
%\end{eqnarray}
\begin{equation}
\tau_{r\text{(e)}}^{-1} = 2 \tau_{r0}^{-1} \sqrt{E_{*}} f_1 (E_{*}).
\label{tau_ra}
\end{equation}

We see that Eqs. (\ref{tau_ra}) and (\ref{tau_rs}) coincide. This is not
surprising because in our model the vertical and horizontal UCN motions
along each main axis are separated. For a rectangular UCN trap the
absorption rate by side walls depends only on the horizontal UCN velocities,
which do not depend on the height $h$ above the trap bottom, as in the
oversimplified formula (\ref{tau_s}). However, as can be seen from Fig. \ref%
{FigTauC} below, a similar coincidence does not hold for arbitrary UCN trap
shapes, where the two horizontal UCN velocity components do not separate.

At $E_{*} \ll 1$ Eqs. (\ref{tau_rg}) and (\ref{tau_rs}) or (\ref{tau_ra})
simplify to 
\begin{equation}
\frac{\tau_{r\text{(g)}}^{-1}}{\tau_{r0}^{-1}} \simeq \frac{4}{5} E_{* },
\quad \frac{ \tau _{r\text{(s)}}^{-1}}{\tau_{r0}^{-1}} = \frac{\tau_{r\text{%
(e)}}^{-1}}{\tau_{r0}^{-1}} \simeq \frac{4}{3}E_{*}.  \label{Taylor}
\end{equation}
These low-energy Taylor expansions 
%for $\tau_{Lg}^{-1}\left( E\right) $ and $\tau_{LS}^{-1}\left( E\right) $ 
differ already in the first linear-term coefficient.

\begin{figure}[tb]
\centering \includegraphics{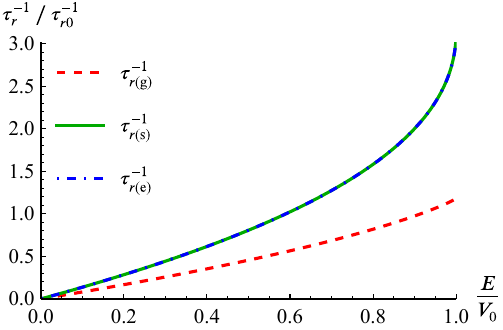}
\caption{The energy dependence of UCN loss rate $\protect\tau_{r}^{-1}$ from
the absorption by side walls only, corresponding to rectangular $L_x \times
L_y$ trap, for the isotropic UCN velocity distribution at trap bottom given 
by Eq. (\ref{f0}) and illustrated in Fig. \ref{FigIsotropic}. 
The calculation is performed in three ways: the standard method,
giving Eq. (\protect\ref{tau_rg}) (dashed red), the oversimplified method,
giving Eq. (\protect\ref{tau_rs}) (solid green), and the proposed
calculation method for a rectangular trap, giving Eq. (\protect\ref{tau_ra})
(dot-dashed blue).}
\label{FigTauR}
\end{figure}

In Fig. \ref{FigTauR} we compare the UCN absorption rates by side walls
given by Eqs.(\ref{tau_rg}),(\ref{tau_rs}) and (\ref{tau_ra}). The result in
Eq. (\ref{tau_rg}), obtained by the standard method, differs strongly from
that in Eqs. (\ref{tau_rs}) and (\ref{tau_ra}), as one can see already from
their Taylor expansions in Eq. (\ref{Taylor}). This difference appears
because the standard method (g), described by Eq. (\ref{tau_rg}) and coming
from Eqs. (\ref{mu(E)})--(\ref{rho}), assumes an isotropic UCN velocity
distribution at any height. This assumption is not fulfilled for isotropic 
UCN velocity distribution at trap bottom given by Eq. (\ref{f0}) because at
large height the vertical velocity component is reduced by gravity, while
the horizontal UCN velocity is not affected by the gravity. Hence, the
normal-to-wall horizontal UCN velocity component, which enters Eq. (\ref{mu}%
), is larger than the one assumed in Eq. (\ref{tau_rg}). Therefore, Eq. (\ref%
{tau_rg}) gives a smaller absorption rate by side walls. This is
especially important for the size extrapolation, which now depends on trap
shapes. On contrary, in Eq. (\ref{tau_ra}) the
isotropic UCN velocity distribution is assumed only at $z=0$, which
corresponds to the averaging in Eq. (\ref{tau(E)}).

\subsection{Absorption by cylindrical side wall}

The number of UCN collision with the side wall of a straight vertical
cylinder of radius $R$ during a long time $t\gg R/v_{xy}$ and the
corresponding absorption rate, in analogy with Eqs. (\ref{N}) and (\ref{tau}%
), are given by: 
\begin{equation}
\mathcal{N}_{c}=\frac{tv\sin \theta }{R\text{crd}\varphi },\quad \tilde{\tau}%
_{c\text{(e)}}^{-1}(\bm{v})=\frac{\mu (v_{\perp })\mathcal{N}_{c}}{t},
\label{tau_c}
\end{equation}%
where $\text{crd}\varphi =2\sin (\varphi /2)$ is the chorde length in a unit
circle, and $v\sin \theta $ is the neutron speed in $xy$ plane (see Fig. \ref%
{cylFig}). The neutron velocity component normal to the cylinder walls is
expressed as 
\begin{equation}
v_{\perp }=v\sin \theta (\tilde{\bm{v}}_{xy}\cdot \tilde{\bm{n}}_{xy})=v\sin
(\varphi /2)\sin \theta ,
\end{equation}%
where $\tilde{\bm{v}}_{xy}=(\cos \varphi -1,\sin \varphi )/\text{crd}\varphi 
$ is the unit direction vector of UCN velocity in the $xy$-plane, $\tilde{%
\bm{n}}=(\cos \varphi ,\sin \varphi )$ is the unit vector normal to circle
in the $xy$-plane at the intersection point.

\begin{figure}[tb]
\centering
\includegraphics[width=0.67\linewidth]{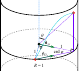}
\caption{Neutron path between the reflections from the side wall of a
cylindrical trap. This scheme also illustrates the notations, used in the
text and formulas.}
\label{cylFig}
\end{figure}

Averaging Eq. (\ref{tau_c}) over angles, we get the absorbtion rate 
\begin{equation}
\tau^{-1}_{c\text{(e)}} = \tau_{c0}^{-1} \sqrt{E_*} \left( \arcsin\sqrt{E_*}
+ \frac{f_1(E_*)}{2} \right),  \label{tau_ca}
\end{equation}
where we introduced $\tau_{c0}^{-1} = \eta v_{\lim} / 2 R$ describing the
UCN absorption rate by cylindrical side wall in the order of magnitude.

The standard calculation method described by Eqs. (\ref{mu(E)})-(\ref{tau_g}%
) for the cylindrical trap of radius $R$ gives the side-wall absorption rate 
\begin{equation}
\tau_{c\text{(g)}}^{-1} \left(E\right) = 3 \tau_{c0}^{-1} \frac{f_2(E_*)}{%
\sqrt{E_*}}.  \label{tau_cg}
\end{equation}
It resembles the expression (\ref{tau_rg}) derived for the rectangular trap
except for the coefficient $\tau_{r0}^{-1}\to \tau_{c0}^{-1}$.

The oversimplifies method neglecting the gravity and described by Eq. (\ref%
{tau_s}) gives 
\begin{equation}
\tau _{c\text{(s)}}^{-1} \left(E\right) =2 \tau_{c0}^{-1} \sqrt{E_{*}} f_1
(E_{*}).  \label{tau_cs}
\end{equation}

The results given by Eqs. (\ref{tau_cs}) and (\ref{tau_ca}) differ, which
follows from the quadratic term of their Taylor expansions at $E_* \ll 1$: 
\begin{equation}
\frac{ \tau_{c\text{(s)}}^{-1} } { \tau_{c0}^{-1} } \simeq \frac{4}{3}E_* +%
\frac{2}{5}E_*^2, \quad \frac{ \tau_{c\text{(e)}}^{-1} } { \tau_{c0}^{-1} }
\simeq \frac{4}{3}E_* +\frac{4}{15}E_*^2.
\end{equation}
Eq. (\ref{tau_cg}), corresponding to the standard calculation method,
differs from two other methods much stronger, already in the linear order: 
\begin{equation}
\frac{ \tau_{c\text{(g)}}^{-1} } { \tau_{c0}^{-1} } \simeq \frac{4}{5}E_*.
\end{equation}

\begin{figure}[tb]
\centering
\includegraphics{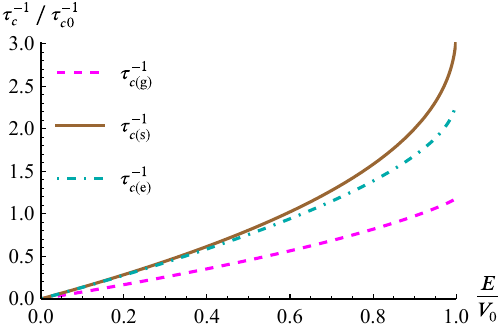}
\caption{The energy dependence of UCN loss rate $\protect\tau_{c}^{-1}$ from
the absorption by side wall of a straight cylindrical trap of radius $R$ 
for the isotropic UCN velocity distribution at trap bottom given by Eq. (\ref{f0}) 
and illustrated in Fig. \ref{FigIsotropic}.
The calculation and averaging over the UCN incidence angle is performed in
three ways: the standard method, giving Eq. (\protect\ref{tau_cg}) and shown
by dashed magenta line, the oversimplified method, resulting to Eq. (\protect
\ref{tau_cs}) (solid brown line), and the proposed calculation method for a
cylindrical trap, giving Eq. (\protect\ref{tau_ca}) (dot-dashed cyan line).}
\label{FigTauC}
\end{figure}

In Fig. \ref{FigTauC} we compare the UCN absorption rates by cylindrical
side walls calculated by all three methods and given by Eqs. (\ref{tau_ca}),(%
\ref{tau_cg}) and (\ref{tau_cs}). We see that the results of proposed and
oversimplified methods, given by Eqs. (\ref{tau_ca}) and (\ref{tau_cs})
correspondingly, now differ but not strongly, mainly at large neutron energy
approaching the potential barrier height $V_0$. On contrary, the UCN loss
rates calculated by the standard method and given by Eq. (\ref{tau_cg}) is
very different. This difference is similar to the case of rectangular side
walls, considered in Sec. \ref{SecRectangularSideWalls} and illustrated in
Fig. \ref{FigTauR}, and have the same origin, discussed in the end of Sec. %
\ref{SecRectangularSideWalls}.

\subsection{Total absorption rate and size extrapolation}

From Figs. \ref{FigTauZ}, \ref{FigTauR}, and \ref{FigTauC} we see that the
UCN absorption rate and its dependence on UCN energy differ strongly for
trap bottom and side walls. This means that the UCN absorption rate changes
differently if the trap dimensions are reduced along the vertical $z$ or
horizontal $x,y$ axes. This is very important because it affects the
procedure of size extrapolation, on which all current precise $\tau_\text{n}$
UCN storage measurements are based to account for the difference $\gtrsim 2$%
\% between the measured and extracted neutron lifetime. Our calculations
show that the result of the size extrapolation depends strongly on the
shapes of large and reduced UCN traps, i.e. on the position and shape of the
trap insert in UCN-$\tau$ experiments.

To illustrate this message and to estimate possible error in the estimates
of neutron loss rate we now compare the UCN absorption rate calculated using
the above three methods for a rectangular (cylindrical) UCN trap of typical
dimensions $L_{x},L_{y} \sim h_{\lim}$ for rectangular and $R \sim h_{\lim}$
for cylindrical traps. Evidently, the total absorption rate $\tau^{-1}$ is
given by the sum of the absorption rate $\tau_{z}^{-1}$ by trap bottom due
to the vertical UCN motion and the absorption rate $\tau_{r}^{-1}$ ($%
\tau_{c}^{-1}$) by side walls due to the horizontal %(longitudinal) 
UCN velocity.

For the rectangular trap the total absorption rate 
\begin{equation}
\tau_\text{R}^{-1} \left(E_*\right) = \tau_{z}^{-1} (E_*) + \tau_{r}^{-1}
(E_*).  \label{tau_full_R}
\end{equation}
It is convenient to introduce the dimensionless parameter $h_{*}=h_{\lim}
(L_{x}+L_{y}) / 2 L_{x} L_{y} = h_{\lim}(L_{x}^{-1}+L_{y}^{-1})/2$, which
describes the UCN trap size and enters all the expressions for absorption
rate. At $L_{x}=L_{y}=L$ this size parameter $h_{* }=h_{\lim}/L$, and $%
\tau_{r0}^{-1} = 2 \tau_{z0}^{-1} h_*$. Combining Eqs. (\ref{tau_full_R}), (%
\ref{tau_zs}) and (\ref{tau_rs}) we obtain for the oversimplified method
without gravity the following total UCN absorption rate: 
\begin{equation}
\tau_{\text{R(s)}}^{-1} \left(E_*\right) = \tau_{z0}^{-1} \frac{f_1(E_{*})}{%
\sqrt{E_{* }}} \left( 1+4h_{* }E_{* } \right).  \label{tauFS}
\end{equation}%
Combining Eqs. (\ref{tau_full_R}), (\ref{tau_zg}) and (\ref{tau_rg}) gives
the result of standard method where the gravity changes only the UCN energy
and concentration as a function of height: 
\begin{equation}
\tau_{\text{R(g)}}^{-1} \left( E_*\right) =\frac{3}{2} \tau_{z0}^{-1} \frac{%
f_1(E_{* })}{\sqrt{E_{* }}} \left( 1+4 h_{* } \frac{f_2(E_{* })} {f_1( E_{*
}) } \right).  \label{tauFg}
\end{equation}%
Eqs. (\ref{tau_full_R}), (\ref{tau_za}) and (\ref{tau_ra}), derived exactly 
for the isotropic velocity distribution (\ref{f0}) at trap bottom, give
the following total UCN absorption rate by the walls of rectangular trap: 
\begin{equation}
\tau_{\text{R(e)}}^{-1} \left( E_*\right) =\tau_{z0}^{-1} \frac{f_1 (E_{*}) 
}{\sqrt{E_{* }}} \left( \frac{\arcsin \sqrt{E_{* }}}{f_1 (E_{*}) }+4 h_{*
}E_{* } \right).  \label{tauF}
\end{equation}

Let us compare a rectangular trap with a cylindrical one at the same volume
to height ratio, i.e. the same base area $V/h_{\lim }=L^{2}=\pi R^{2}$. Then 
$h_{\ast }=h_{\lim }/(R\sqrt{\pi })$ and $\tau _{c0}^{-1}=\sqrt{\pi }\tau
_{z0}^{-1}h_{\ast }$. Performing the same steps as for a rectangular trap,
but using 
\begin{equation}
\tau _{\text{C}}^{-1}\left( E_{\ast }\right) =\tau _{z}^{-1}(E_{\ast })+\tau
_{c}^{-1}(E_{\ast })  \label{tau_full_C}
\end{equation}%
instead of Eq. (\ref{tau_full_R}), we get the analytical formulas for the
total UCN absorption rate in a cylindrical trap for all three methods.
Combining Eqs. (\ref{tau_full_C}), (\ref{tau_zs}) and (\ref{tau_cs}) we
obtain the oversimplified result 
\begin{equation}
\tau _{\text{C(s)}}^{-1}\left( E_{\ast }\right) =\tau _{z0}^{-1}\frac{%
f_{1}(E_{\ast })}{\sqrt{E_{\ast }}}\left( 1+2\sqrt{\pi }h_{\ast }E_{\ast
}\right) ,  \label{tauFSc}
\end{equation}%
Combining Eqs. (\ref{tau_full_C}), (\ref{tau_zg}) and (\ref{tau_cg}) gives
the standard-method estimate of UCN absorption rate 
\begin{equation}
\tau _{\text{C(g)}}^{-1}\left( E_{\ast }\right) =\frac{3}{2}\tau _{z0}^{-1}%
\frac{f_{1}(E_{\ast })}{\sqrt{E_{\ast }}}\left( 1+2\sqrt{\pi }h_{\ast }\frac{%
f_{2}(E_{\ast })}{f_{1}(E_{\ast })}\right) ,  \label{tauFGc}
\end{equation}%
Eqs. (\ref{tau_full_C}), (\ref{tau_za}) and (\ref{tau_ca}), based on Eqs. (\ref{tau})-(\ref{f0}), give 
\begin{eqnarray}
\tau _{\text{C(e)}}^{-1}\left( E_{\ast }\right) &=&\tau _{z0}^{-1}\Bigg[%
\frac{\arcsin \sqrt{E_{\ast }}}{\sqrt{E_{\ast }}}\left( 1+\sqrt{\pi }h_{\ast
}E\right)  \notag \\
&&+\frac{\sqrt{\pi }}{2}h_{\ast }\sqrt{E_{\ast }}f_{1}(E)\Bigg].
\label{tauFc}
\end{eqnarray}

\begin{figure}[tb]
\label{FigTau}\centering
\includegraphics{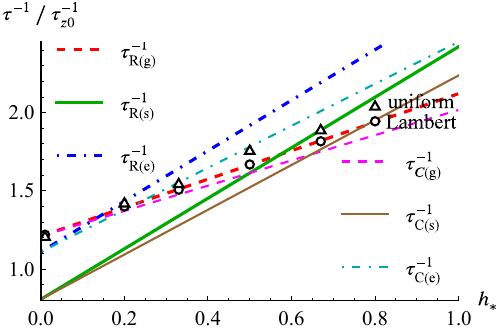}
\caption{The dependence of UCN loss rate $\protect\tau^{-1}$ on the reduced
inverse trap size $h_{*}$ at $E_{* }=1/2$ for the
isotropic UCN velocity distribution at trap bottom given by Eq. (\ref{f0}) 
and illustrated in Fig. \ref{FigIsotropic}. Rectangular and cylindrical traps
are taken with equal areas, so that $L=R\protect\sqrt{\protect\pi}$. The
averaging over the UCN incidence angle is performed in three ways: (i) the
standard method, resulting to Eqs. (\ref{tauFg}) and (\ref%
{tauFGc}) and illustrated by dashed lines (the red line corresponds to
rectangular trap and the magenta line corresponds to cylindrical trap),
function of height but not its angular distribution, (ii) the
oversimplified method, neglecting all gravity effects and giving Eqs. (%
\ref{tauFS}) and (\ref{tauFSc}) (green and brown solid
lines), and (iii) the enhanced calculation method resulting to Eqs. 
(\ref{tauF}) and (\ref{tauFc}), shown by the blue and cyan dot-dashed
lines. The circle and triangle symbols give the results of our Monte-Carlo 
simulations in rectangular traps of the same dimensions including 
the diffuse UCN scattering by the trap walls with 
probability $p_\text{d}=0.1$ and obeying the Lambert's law (circles) and 
uniform distribution (triangles) of diffusively scattered neutrons, 
as discussed in Sec. \ref{SecMonteCarlo}. }
\end{figure}

In Fig. \ref{FigTau} we compare the geometrical size scaling of UCN loss
rates in a rectangular and cylindrical traps calculated by three different
methods for a typical UCN energy $E=V_0/2$. The green and brown solid lines
show the oversimplified result in Eqs. (\ref{tauFS}) and (\ref{tauFSc}) for
rectangular and cylindrical traps correspondingly, where the gravity effects
are neglected and the isotropic velocity distribution of UCN is assumed. The
red and magenta dashed lines illustrate the improved approximate formulas (%
\ref{tau_g}) and (\ref{mu(E)}), applied in Refs. \cite%
{Serebrov2008PhysRevC.78.035505,Serebrov2018PhysRevC.97.055503} and
resulting to Eqs. (\ref{tauFg}) and (\ref{tauFGc}), where the gravity effect
is included via the height-dependent UCN energy and concentration, but the
isotropic velocity distribution of UCN is assumed at any height and the
collision rate with trap bottom is proportional to UCN vertical velocity.
The dot-dashed blue and cyan lines show the UCN absorption rates by
rectangular and cylindrical trap walls given by Eqs. (\ref{tauF}) and (\ref%
{tauFc}), calculated exactly for the isotropic velocity distribution 
(\ref{f0}) of UCN at trap bottom.

As one can see from Fig. \ref{FigTau} or from Eqs. (\ref{tauFg})-(\ref{tauFc}%
), the absorption rate $\tau^{-1}$ depends linearly on $h_*$. All three
calculation methods give very different values $\tau^{-1}(h_*=0)$,
corresponding to the absorption rate by trap bottom only and describing a
very wide UCN trap. This difference exceeds $10$\% even for the improved
approximate method including gravity, which may result to the error up to $%
\sim 2$ seconds in the estimate of neutron lifetime $\tau_n$. Hence, the
size extrapolation by changing the trap base area gives very different UCN
absorption rates for the considered three methods.

\begin{figure}[tb]
\centering
\includegraphics{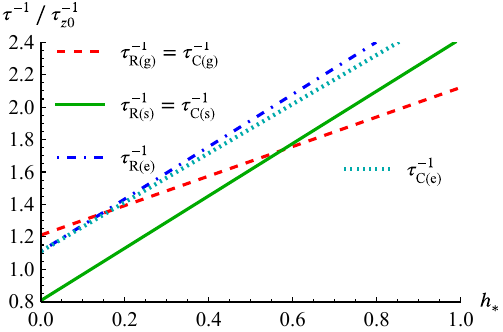}
\caption{ The dependence of UCN loss rate $\protect\tau^{-1}$ on the reduced
inverse trap size $h_{*}$ at $E_{* }=1/2$. The difference from Fig. \protect
\ref{FigTau} is that here the comparison is at $L=2R$. }
\label{FigTau2}
\end{figure}

The linear $\tau ^{-1}(h_{\ast })$ dependence and the same value $\tau
^{-1}(h_{\ast }=0)$ for the rectangular and cylindrical traps calculated by
the same method can be used to scale the plots in Fig. \ref{FigTau} by
changing the definition of $h_{\ast }$ for the cylindrical trap from $%
h_{\ast }=h_{\lim }/(R\sqrt{\pi })$ to $h_{\ast }=h_{\lim }/(2R)$,
corresponding to $L=2R$ and $\tau _{c0}^{-1}=2\tau _{z0}^{-1}h_{\ast }$. In
this case, the standard (g) and oversimplified (s) formulas for cylindrical trap
coincide exactly with those obtained for a rectangular trap and given by
Eqs. (\ref{tauFg}), (\ref{tauFS}). This is illustrated in Fig. \ref{FigTau2}. 
Only the isotropic UCN velocity distribution (\ref{f0}), resulting to Eqs. 
(\ref{tauF}) and (\ref{tauFc}), gives 
\begin{equation}
	\tau _{\text{C(e)}}^{-1}(E_{\ast })=\tau _{z0}^{-1}\!\left[ \frac{\arcsin \!%
		\sqrt{E_{\ast }}}{\sqrt{E_{\ast }}}\left( 1+2h_{\ast }E\right) +h_{\ast }\!%
	\sqrt{E_{\ast }}f_{1}(E)\right] \!,  \label{tauFcc}
\end{equation}
i.e. slightly different results for cylindrical
and rectangular traps with the deviation less than 4\%, see Fig. \ref%
{FigTau2}, because it explicitly takes into account the specular reflection
of neutrons from the cylindrical walls. % \begin{equation}
% 	\tau_{\text{C(e)}}^{-1} \left( E_*\right) =\tau_{z0}^{-1}
% 	\left( 
% 		\frac{\arcsin \sqrt{E_{* }}}{\sqrt{E_{* }}}+8 h_{*} E_{*} f_3(E_{*})
% 	\right).  \label{tauFcc}
% \end{equation}
% \begin{multline}
% 	\tau_{\text{C(e)}}^{-1} \left( E_*\right) =\tau_{z0}^{-1}
% 	\Bigg[ 
% 		\frac{\arcsin \sqrt{E_{* }}}{\sqrt{E_{* }}} \left( 1 + 2 h_* E \right) \\
% 		+ h_* \sqrt{E_*} f_1(E)
% 	\Bigg].  \label{tauFcc}
% \end{multline}

%All results are illustrated in Figs. \ref{FigTau} and \ref{FigTau2}.

\section{Monte-Carlo simulation and the role of diffuse scattering}
\label{SecMonteCarlo}

Before we considered only the specular UCN reflection from the trap walls.
However, there is a small probability $p_{\text{d}} \lesssim 0.1$ of a
diffuse UCN reflection by an arbitrary angle, which conserves only the
absolute value of UCN velocity and its energy. This elastic diffuse
reflection is much more probable than the UCN inelastic scattering or
absorption, $p_{\text{d}} \gg \eta $, and must be considered. The rare
diffuse scatterings mean that the UCN velocity direction in Eq. (\ref{Pa})
after many reflections may becomes arbitrary. As we show below, the diffuse 
scattering changes the uniform and isotropic number distribution function 
(\ref{f0}) of UCN velocity at trap bottom, making instead the UCN density 
distribution $n( h,\bm{v}) $ to be isotropic.  

Usually, in Monte-Carlo simulations one takes\cite%
{Fomin2023,Fomin2019,Ayres2018,Fomin2018,Fomin2017,Serebrov2013MC} the
Lambert's cosine law for UCN angular velocity distribution after the diffuse
scattering :%
\begin{equation}
p_\text{L}  \left(\theta \right)=2\cos \theta ,  \label{Lambert}
\end{equation}%
where $\theta $ is the angle between the normal to the wall and the UCN
velocity after the diffuse scattering. This distribution $ p (\theta )$ is normalized: 
\begin{equation}
\int_0^{\pi /2} d\theta \, p \left(\theta \right) \sin\theta =1.
\label{Norm}
\end{equation}% 
Even in optics, the Lambert's law is not
universal, and there are many deviations from it \cite{LambertReview2020,
Hecht1976, Mamouei2021}. These deviations are very important in optical
spectroscopy, for example, for biological analysis \cite{Mamouei2021}. For
diffuse neutron scattering, there is no strict derivation of the Lambert's
law at all. Small defects in the trap walls, having a characteristic size
smaller than the UCN wavelength $\lambda \sim 100$ nm, give a uniform
angular distribution $I(\theta )=\const$ of the probability of UCN diffuse
scattering, quite different from the Lambert's law (\ref{Lambert}). Such
defects include surface roughness, nanopores, impurities, etc. Even
ultrasmooth surfaces obtained after their multiple processing always have
roughness on the submicron scale of $\lesssim 100$ nm, clearly visible with
an atomic force and/or scanning electron microscope. For example, the
ultrasmooth surface of nickel or beryllium intended for ultraviolet mirrors
of space satellites has noticeable roughness of $\lesssim 100$ nm in size,
visible with an atomic force microscope \cite{Chkhalo2019}. Beryllium was
previously actively used for UCN traps \cite{Ignatovich/1990,Ignatovich1996}.
 Modern UCN material traps use a coating of perfluoropolyether -- Fomblin
brand oil, which does not contain hydrogen \cite%
{Serebrov2008PhysRevC.78.035505,Serebrov2018PhysRevC.97.055503}, the surface
of which at the submicron scale $\lesssim 100$ nm also has strong roughness 
\cite{Russell2007,Ganesh2014,Masciullo2018}. Large wall roughness or pores
of size $d\gg \lambda $ in the material trap, which scatter with a small
transmission of the wave vector $\sim 1/d$, blur the peak of specular
reflection in the angular distribution of the UCN velocity after diffuse
scattering and also do not give the Lambert's law. The main argument in favor
of Lambert's law is the principle of detailed balance of the UCN velocity
distribution, which requires that the probability of scattering from walls
is proportional to $\cos \theta $, since the collision frequency is also $%
\propto \cos \theta $ (see p. 96 of \cite{Ignatovich/1990}). Neglecting
gravity, the Lambert's law indeed maintains an isotropic velocity distribution
and even makes an anisotropic gas more isotropic, changing the angular
distribution of neutron velocities at the wall surface to their isotropic
distribution. 
The Lambert's law provides an isotropic velocity distribution only if gravity is 
negligible. In the large open-top UCN material traps used in neutron lifetime measurements 
\cite{Serebrov2008PhysRevC.78.035505,Serebrov2018PhysRevC.97.055503}, the velocity isotropy 
is partially provided by specular reflections from curved walls with various local surface orientations. 

The microscopic calculation of the UCN diffuse scattering law is a complicated and still open 
problem. Recently, a simple experimental test to measure the deviations from the Lambert's 
cosine law for UCN diffuse scattering has been proposed \cite{Grigoriev2024Lambert}, but such 
an experiment is not yet performed. For our Monte-Carlo calculations we use two limiting 
cases of the generalized UCN diffuse scattering law 
\begin{equation}
	p_\text{d}  (\theta )=p_\text{L}\, 2\cos \theta +1-p_\text{L},  \label{LambertD}
\end{equation}%
where $p_L$ is the probability that the diffuse scattering obeys the Lambert's cosine law, 
and $1-p_L$ that it obeys the isotropic diffuse reflection. Evidently, this angular 
distribution (\ref{LambertD}) also satisfies the normalization condition (\ref{Norm}).

The results of our Monte-Carlo calculations of the UCN absorption rate in rectangular 
UCN traps of various bottom area for the Lambert's cosine law, $p_\text{L}=1$, and for isotropic 
law, $p_\text{L}=0$, of diffuse neutron scattering by trap walls are shown in Fig. \ref{FigTau} 
by open circles and triangles correspondingly. The calculations with Lambert's diffuse 
scattering confirm the result (g) given by Eq. (\ref{tauFg}), obtained using the standard 
method described by Eq. (\ref{tau_g}). On contrary, the result (e) in Eq. (\ref{tauF}), 
obtained assuming the uniform UCN distribution (\ref{f0}), is not confirmed by our 
Monte-Carlo calculations. Let us understand the physical reason of these results.

If the diffuse scattering kept the overall isotropic velocity distribution (\ref{f0}), 
one could use Eqs. (\ref{tau(E)}) with the absorption rate given by the exact Eq. (\ref{tau}). 
This would confirm the analytical results (e) obtained in Sec. \ref{SecCalulations}. However, 
this simple approach does not work because the diffuse scattering violates the uniform velocity 
distribution (\ref{f0}). Instead, as expected from the principle of detailed balance of the UCN 
velocity distribution \cite{Ignatovich/1990}, it is the UCN density near the wall rather than 
the UCN number has the uniform velocity distribution due to the Lambert's diffuse scattering, 
as we illustrate below by our Monte-Carlo calculations.

\iffalse
To take into account the probability $p_{\text{d}}$ of diffuse UCN
scattering, we average Eq. (\ref{tau}) over the velocity directions taking
their isotropic distribution: 
\begin{equation}
\tau _{\text{(e)}}^{-1}\left( E\right) =\int \frac{d\varOmega}{4\pi }\tilde{%
\tau}_{\text{(e)}}^{-1}(\bm{v}).
\end{equation}%
Eq. (\ref{tau(E)}) does not contain the kinematic factor $\cos \theta $ as
in the calculation of collision rate because it is already contained in the
absorption rate $\tau _{\text{(e)}}^{-1}$. As the neutrons reflect
specularly many times before their diffusive scattering, Eqs. (\ref{Pa}) and
(\ref{tau}) remain valid between the diffuse reflections, and the
replacement $\tilde{\tau}_{\text{(e)}}^{-1}(\bm{v})\rightarrow \tau _{%
\text{(e)}}^{-1}(E)$ accounts for rare diffuse reflections. As we discuss in
Sec. \ref{SecDiscussion}, the diffuse elastic UCN scattering changes their
velocity distribution depending on the trap shape. However, the UCN velocity
distribution remains isotropic on a height $\bar{h}_{c}\ll h_{\max }$,
substantiating Eq. (\ref{tau(E)}).
\fi

\begin{figure}[tb] \label{Fig_MC_w}
    \centering

    \includegraphics{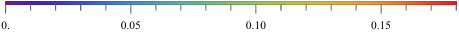} \\
    \includegraphics[height=0.325\linewidth]{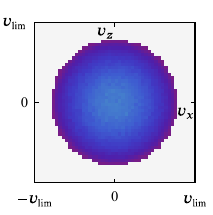}
    \includegraphics[height=0.32\linewidth]{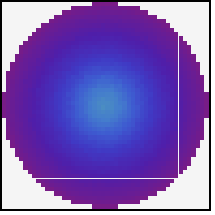}
    \includegraphics[height=0.32\linewidth]{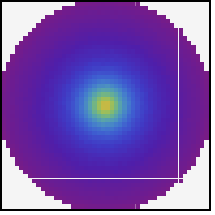} \\

    \includegraphics[height=0.32\linewidth]{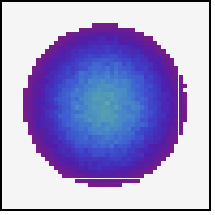}
    \includegraphics[height=0.32\linewidth]{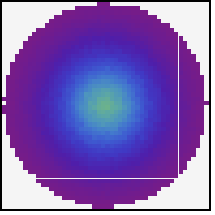}
    \includegraphics[height=0.32\linewidth]{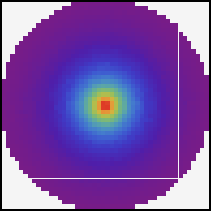} \\

    \includegraphics[height=0.32\linewidth]{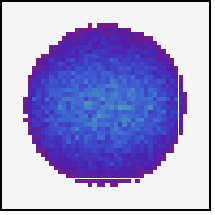}
    \includegraphics[height=0.32\linewidth]{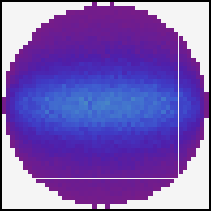}
    \includegraphics[height=0.32\linewidth]{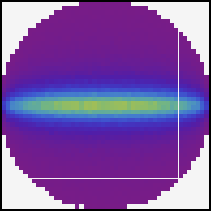}
    \caption{Slices of neutron probability (density) distribution (PDF) $n (v_x,v_z)$ at 
	height $h=0.5$ m (left column), 
    $h=0.1$ m (center column) and $h=0.01$ m (right column) in a trap with 
	$1 \times 1$ m$^2$ bottom size. Probability of diffuse (Lambertian)
	scattering $p_\text{d} = p_\text{L} =$ 0.1, 0.0001 and 0 from top to bottom. 
    Data for averaging the PDF are collected within a 1000 s time interval after the 
	thermodynamic equilibrium has been established. The time to establish equilibrium is estimated as
	$\tau_\text{eq} \approx 10 L / (p_\text{d} v_{\lim})$, where 
    $L = \sqrt[3]{L_x L_y h_{\lim}}$. In the case of $p_\text{d}=0$ the equilibrium is never achieved,
    so there is no point to wait for it and we set $\tau_\text{eq}=0$.
    Angular velocity distribution of UCN density is always isotropic and $n (\theta) = \sin (\theta )/ 2$ 
    for $p_\text{d} \neq 0$.
    }
\end{figure}

\begin{figure}[tb]
    \centering
	\includegraphics[height=0.325\linewidth]{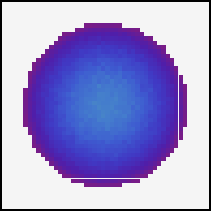}
	\includegraphics[height=0.325\linewidth]{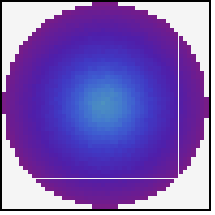}
    \includegraphics[height=0.325\linewidth]{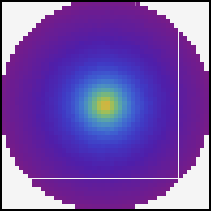}
    \caption{ Same as top row of Fig. \ref{Fig_MC_w} with $p_\text{d} =  0.1$ and $p_\text{L} = 1$, 
	but bottom size is $10000 \times 10000$ m$^2$. The distribution did not 
	change with increasing trap size.
    }
    \label{Fig_MC_wide}
\end{figure}

\begin{figure}[tb]
    \centering

	\includegraphics[height=0.5\linewidth]{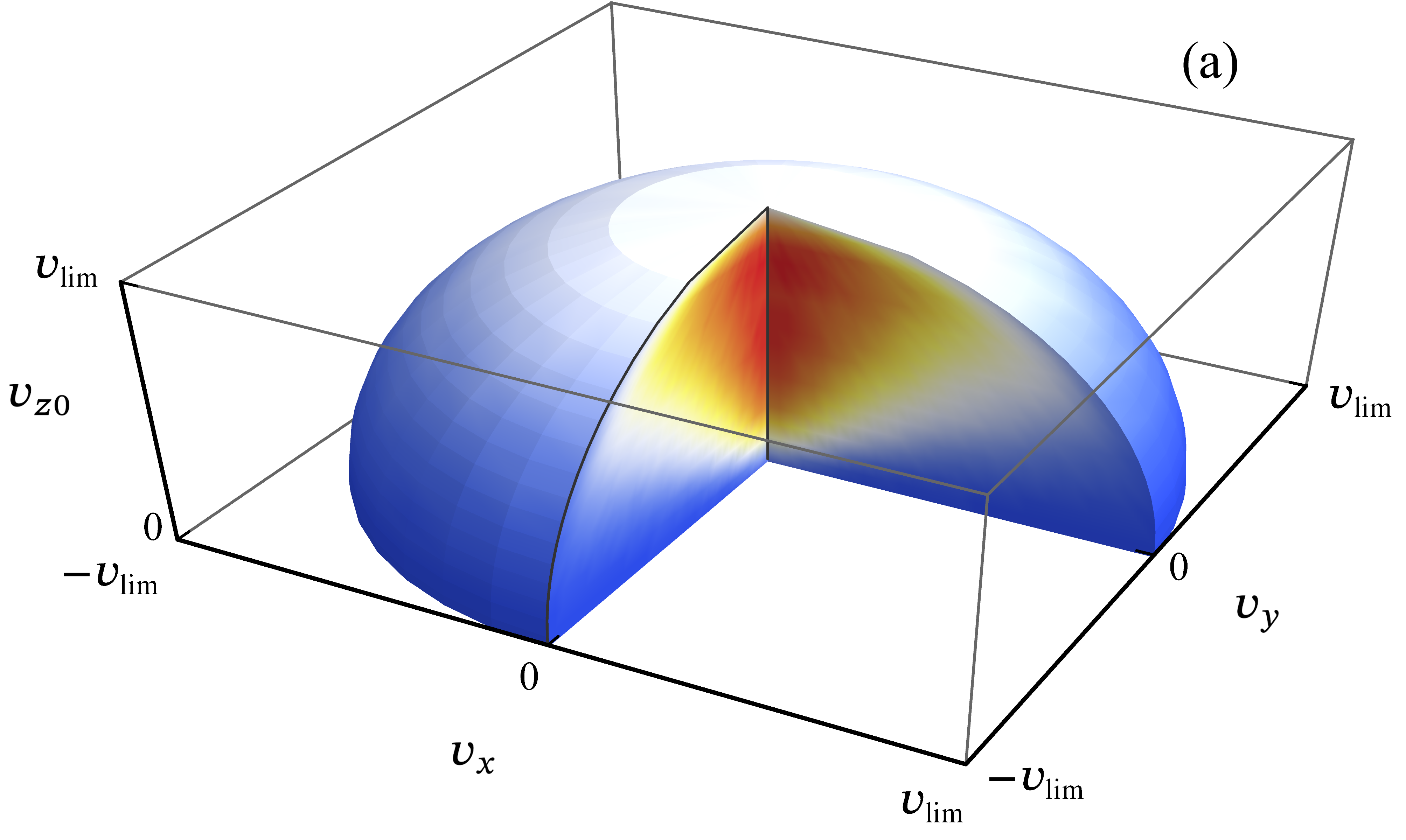}
	\includegraphics[height=0.5\linewidth]{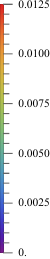}\\

    \includegraphics[height=0.333\linewidth]{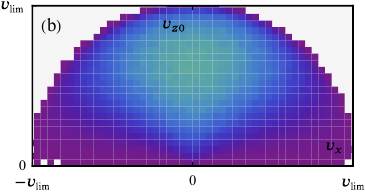}
	\includegraphics[height=0.333\linewidth]{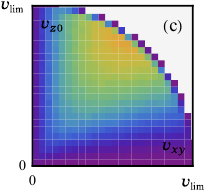} \\
	\includegraphics{legend.pdf}
    \caption{
	(a) Slice of neutron PDF $n (v_x,x_y,v_{z0})$,
	where $v_{z0}=\sqrt{v_z^2+2gz}$ is the vertical speed of a neutron that it has when 
	reaches the trap bottom in the Earth gravity potential.
	(b,c) Neutron PDFs $n (v_x,v_{z0})$ and $n (v_{xy},v_{z0})$, where $v_{xy} = \sqrt{v_x^2+v_y^2}$.
    The snapshot of the distribution was taken at the end of the Monte Carlo simulation, long after 
	thermodynamic equilibrium had been reached.
	One can check numerically that these plots correspond to the distribution by azimuth angle 
	$n(\theta_0) = 3/2 \sin^3 (2\theta_0) \cos (\theta_0)$, where $\theta_0=\arccos(v_{z0}/v_0)$.
	Entering polar coordinates in the $v_x$-$v_{z0}$ plane with polar angle 
	$\alpha=\operatorname{atan2}(v_{z0},v_x)=\arctan (v_{z0}/v_x)$ one can also numerically check that the distribution 
	by this angle is given by $n(\alpha) = (7/11) \sin^2 \alpha$. This distribution is
	symmetric with respect to $\alpha = \pi/2$, which is clear from panel (b).
    } 
    \label{Fig3D}
\end{figure}

In Fig. \ref{Fig_MC_w} we plot the $v_x$-$v_z$ velocity distribution of neutron density on the 
height $h=0.5 h_{\max}$ (left column), $h=0.1$ m (center column) and $h=0.01$ m (right column) 
in a rectangular trap with $1 \times 1$ m$^2$ bottom size for the Lambert's law of diffuse 
scattering with probability $p_\text{d}=0.1$ (upper row), $p_\text{d}=0.0001$ (middle row), and  
$p_\text{d}=0$ 
(lowest row). The circle radius of UCN velocity distribution is reduced with height because 
the UCN kinetic energy and velocity absolute value decrease with height. From Fig. \ref{Fig_MC_w} 
we see that even for a very small probability $p_\text{d}=0.0001$ of diffuse scattering (middle row) 
obeying the Lambert's law the velocity distribution of UCN density becomes isotropic at any 
height after some time exceeding $\tau_\text{eq} \approx 10 L / (p_\text{d} v_{\lim})$ -- the 
time when UCNs reach an equilibrium distribution. This isotropy is somehow expected 
\cite{Ignatovich/1990} because the Lambert's law keeps the detailed balance of the UCN 
velocity distribution. The surprising fact is that this isotropy appears at any height 
when the diffuse scattering is included. This equilibrium isotropic velocity distribution 
of UCN density at any height is not only a consequence of the side-wall diffuse scattering 
but appears even if the scattering by trap bottom is dominating. To show this fact, in 
Fig. \ref{Fig_MC_wide} we plot the velocity distribution of UCN density for $p_\text{d}=0.1$ and 
$p_\text{L}=1$ in a very wide trap $10000 \times 10000$ m$^2$ where the neutrons scatter mainly by trap bottom.  

As we explicitly calculated in Sec. \ref{SubSecDistr}, the isotropic velocity distribution 
of UCN density, obtained in our Monte-Carlo simulations and illustrated in Figs. \ref{Fig_MC_w} 
and \ref{Fig_MC_wide}, means anisotropic velocity distribution of UCN total number, and vice 
versa. In Fig. \ref{Fig3D} we show the 3D color plot of UCN number velocity distribution at the 
trap bottom with the Lambert's law of diffuse scattering with $p_\text{d}=0.1$. We see that it 
differ strongly from the original uniform velocity distribution shown in Fig. \ref{FigIsotropic}. 
This means that the uniform velocity distribution (\ref{f0}) does not hold with the diffuse 
scattering by trap walls even if the probability of this diffuse scattering $p_\text{d} \ll 1$. Hence, 
method (e), although being exact for the UCN distribution (\ref{f0}), does not describe well 
the UCN traps. On contrary, the standard method (g) assumes a different UCN velocity 
distribution, which is much closer to the UCN distribution at equilibrium, illustrated in 
Figs. \ref{Fig_MC_w}-\ref{Fig3D}. Hence, it better describes the UCN absorption rate, as one 
sees from the comparison shown in Fig. \ref{FigTau}. 

If the diffuse UCN scattering does not obey the Lambert's cosine law, e.g. if $p_\text{L}\neq 1$ in 
Eq. (\ref{LambertD}), the velocity distribution of UCN density is not isotropic in equilibrium. 
Then, as we show in Fig. \ref{FigTau},  there are considerable deviations of the UCN absorption 
rate from that calculated using the standard method (g) based on Eq. (\ref{tau_g}) or using the MC 
simulations \cite{Fomin2023,Fomin2019,Ayres2018,Fomin2018,Fomin2017,Serebrov2013MC} assuming the 
Lambert's diffuse scattering. Hence, the knowledge of correct velocity distribution of UCN after 
the diffuse scattering by trap walls is crucial for increasing the accuracy of the estimates of 
UCN loss rate in material traps. A microscopic theoretical model of UCN diffuse scattering by 
material walls is not developed yet, but a simple experimental test of the Lambert's cosine law was 
proposed recently \cite{Grigoriev2024Lambert}.  

We also numerically checked Eq. (\ref{rho}), used in the standard calculations, 
and found that it is valid up to the coefficient $g(E)$, which depends on the UCN energy distribution:
$\rho (E,h) \propto g(E)\sqrt{( E-h^{\prime }) /E}$. However, this additional coefficient $g(E)$ does not affect the energy dependence of the absorption rate in Eq. (\ref{tau_g}) because it is contained both in the numerator and denominator of Eq. (\ref{tau_g}) and cancels.  

\section{Discussion and Conclusions}

\label{SecDiscussion}

In this paper we reanalyze the standard calculation methods of UCN
absorption rate by the walls of material traps, which is crucial for the
accuracy of neutron lifetime measurements. The standard analytical formulas (%
\ref{mu(E)})--(\ref{rho}) take the gravity into account but assume (i) an
isotropic UCN velocity distribution at any height and (ii) the similar
dependence of the effective collision rate on the vertical and horizontal
UCN velocity. The latter evidently contradicts Eq. (\ref{N}) giving the
exact collision rates for rectangular UCN traps, where the vertical and
horizontal neutron motion separate. To analyze how well these approximations of
the standard method are fulfilled and how they affect the estimates 
of neutron loss rate we calculate
the UCN absorption rate by the rectangular and cylindrical trap walls
without these assumptions, i.e. using Eqs. (\ref{N}) and (\ref{mu})
for isotropic and uniform UCN velocity distribution given by Eq. (\ref{f0}). 
Then we compare the analytical results obtained by various methods 
with numerical Monte-Carlo calculations.

We succeeded to perform analytical calculations for UCN absorption rates in 
UCN traps of simple rectangular and cylinder shape in three different models: 
(1) the oversimplified method,
neglecting gravity at all and based on Eqs. (\ref{g}) and (\ref{tau_s}); (2)
the improved formulas (\ref{mu(E)})--(\ref{rho}), applied in
Refs. \cite{Serebrov2008PhysRevC.78.035505,Serebrov2018PhysRevC.97.055503}
as a standard method, where the gravity effect is included via the
height-dependent UCN energy and concentration only; (3) by a direct
calculation using Eqs. (\ref{mu}),(\ref{N}) and (\ref{tau}), which give Eq.(\ref{tau(E)}) 
for the uniform UCN velocity distribution (\ref{f0}) illustrated in Fig. \ref{FigIsotropic}. 
%The calculated neutron absorption rate depends much stronger on the calculation method than on the trap shape, rectangular or cylindrical. 

Our results of Sec. \ref{SecCalulations}, illustrated in Figs. \ref{FigTauZ}-\ref{FigTau2}, 
give several observations. First, the UCN absorption rates calculated by these three methods differ
considerably, by $\gtrsim 10$\%, both for rectangular and cylindrical traps.
It is important because may give an error up to few seconds in the extracted
neutron lifetime $\tau_\text{n}$. 
Second, the results of size extrapolation depend strongly on the trap shape
and, hence, must be done with a great care. The size scaling and extrapolation 
to an infinite trap is a standard and necessary procedure for extracting $\tau _\text{n}$. 
We have shown that the change of trap dimensions along the vertical and horizontal directions
affects the UCN loss rate $\tau^{-1}$ differently. Hence, the results of
size extrapolation in UCN-$\tau$ experiments depend on the shape
and size of both trap and its insert. 
Third, in a gravity potential there is a strong difference between the velocity 
distribution of particle number and of particle density. The method (g) described 
by Eq. (\ref{tau_g}) corresponds to the isotropic velocity distribution of UCN density, 
while the method (e) corresponds to the isotropic and uniform velocity distribution 
of UCN number at trap bottom given by Eq. (\ref{f0}). 
Forth, we obtain explicit analytical formulas for the UCN absorption rates in the 
material traps of simplest geometry, rectangular and cylindrical, by both methods (g) and (e), 
as well as by the oversimplified method (s) neglecting gravity.

Our numerical Monte-Carlo calculations in Sec. \ref{SecMonteCarlo} also give 
several important results. First, they show that 
with UCN diffuse scattering by trap walls according to the Lambert's cosine law 
the UCN absorption rate is quite accurately described by the analytical formulas 
corresponding to the standard method (g), given by Eq. (\ref{tau_g}) and used 
in Refs. \cite{Serebrov2008PhysRevC.78.035505,Serebrov2018PhysRevC.97.055503}, 
as illustrated in Fig. \ref{FigTau}. 
This result means that the diffuse scattering makes the UCN density rather than 
the UCN number at trap bottom to have an isotropic velocity distribution. This 
conclusion is directly confirmed by our Monte-Carlo calculations, as illustrated 
in Figs. \ref{Fig_MC_w}-\ref{Fig3D}. Second, the UCN absorption rate depends 
considerably on the law of diffuse neutron scattering by material walls, 
as illustrated in Fig. \ref{FigTau}. For an isotropic diffuse neutron scattering by 
trap walls the UCN velocity distribution differs essentially from that for Lambert's 
cosine law (compare the open triangles and circles in Fig. \ref{FigTau}). 
 Hence, any deviations from the Lambert's 
cosine law of diffuse UCN scattering may considerably affect the estimates 
of UCN losses in material traps and the accuracy of neutron lifetime extracted 
from the measurements. These deviations may be partially 
responsible for the four-second 
difference between the UCN-$\tau$ measurements using material 
\cite{Serebrov2018PhysRevC.97.055503} and magnetic \cite{GONZALEZALONSO2019165} traps.

A possible way to increase the accuracy of UCN loss-rate estimates is to use
the Monte-Carlo simulations \cite%
{Fomin2023,Fomin2019,Ayres2018,Fomin2018,Fomin2017,Serebrov2013MC}, where
the trajectory of each neutron is calculated with allowance for gravity.
These simulations also confirm \cite%
{Fomin2023,Fomin2019,Fomin2018,Fomin2017,Serebrov2018PhysRevC.97.055503} the
standard method of UCN loss calculation given by formulas (\ref{mu(E)})--(\ref{rho}) and used to
analyze the experimental data in Refs. \cite%
{Serebrov2008PhysRevC.78.035505,Serebrov2018PhysRevC.97.055503}. 
This method and the corresponding our analytical formulas are marked by symbol (g). 
However, the physical model used in these numerical calculations requires some
clarification. In the Monte-Carlo simulations, usually, the Lambert's cosine
law of the angular dependence of UCN diffuse scattering intensity is applied 
\cite{Fomin2023,Fomin2019,Ayres2018,Fomin2018,Fomin2017,Serebrov2013MC}. 
In Sec. \ref{SecMonteCarlo}, between Eqs. (\ref{Norm}) and (\ref{LambertD}), 
we argue that the Lambert's law may not hold for the diffuse UCN scattering and 
requires an experimental test, e.g., as proposed in Ref. \cite{Grigoriev2024Lambert}.
\iffalse
Even in optics this Lambert's cosine law is not universal and there are many
deviations from it \cite{LambertReview2020,Hecht1976,Mamouei2021}, which are
quite important in the optical spectroscopy, e.g., used for biological
analysis \cite{Mamouei2021}. For the diffuse neutron scattering there is no
strict derivation of the Lambert's cosine law at all. The short-range
defects in the trap walls, having the characteristic size smaller than the
UCN wave length $\lambda \sim 100\,$nm, give the uniform angular
distribution of diffusive UCN scattering probability rather than the
Lambert's cosine law. These short-range wall imperfections include the
surface roughness, nanopores, impurities, etc. The long-range wall roughness
or large pores of size $d\gg \lambda $ in the trap material, which scatter
with a small wave-vector transfer $\sim 1/d$, smear the specular reflection
peak in the angular distribution of UCN velocity after the diffuse
scattering but do not give the Lambert's cosine law either. The main
argument in favor of the Lambert's cosine law is the principle of detailed
balance of UCN angular distribution, which requires the scattering
probability from the walls to be proportional to $\cos\theta $ because the
collision rate is also $\propto \cos\theta $ (see page 96 of Ref. \cite%
{Ignatovich/1990}). Without gravity the Lambert's cosine law maintains an
isotropic velocity distribution in an isotropic gas and even tends to render
a non-isotropic gas isotropic. 
\fi
The angular distribution of UCN velocity
after the diffuse scattering by a real trap wall is an interesting
fundamental problem, which is beyond the scope of this paper. However, the
estimates of UCN loss probability and the corresponding accuracy of $\tau_%
\text{n}$-measurements depend strongly on this issue, as we illustrated in 
Fig. \ref{FigTau} using our Monte-Carlo simulations. 
%Eq. (\ref{tauF}) for the absorption rate differs from
%that described by the models assuming the isotropic velocity distribution ,
%i.e. (ii) Eqs. (\ref{g}) or (\ref{tauSS}) and (\ref{mu(v)}), or (iii) Eqs. (%
%\ref{tau_g}) or (\ref{tauS}) and (\ref{mu(v)}). Especially, this difference is
%important for the size extrapolation during the extracting the neutron $%
%\beta $-decay lifetime $\tau _{n}$. Further quantitative analysis for the
%particular UCN\ trap size is required to estimate possible errors in the 
%$\tau _{n}$-measurements is a isotropic UCN\ velocity is assumed.

To summarize, we discuss the approximations and possible errors of the
standard calculations of neutron loss rate due to the absorption by trap
walls, which are very important for the precise measurements of neutron
lifetime and, possibly, of its electric dipole moment. To illustrate the
effect of these approximations we calculate analytically and numerically 
the neutron absorption rate by
the walls of rectangular and cylindrical UCN traps using different
methods. Our results show that the standard calculation method of UCN
absorption rate, given by Eq. (\ref{tau_g}) and used in Ref. \cite{Serebrov2018PhysRevC.97.055503}, 
is quite precise if the Lambert's cosine law of neutron scattering by material walls holds. 
However, this method of calculation of UCN losses, as well as the numerous Monte-Carlo calculations \cite%
{Fomin2023,Fomin2019,Fomin2018,Fomin2017,Serebrov2018PhysRevC.97.055503}
based on the Lambert's cosine law of diffuse neutron scattering by trap walls, 
may give a considerable errors when there are deviations from the Lambert's 
cosine law of diffuse UCN scattering by real material walls.
This may partially explain a four-second discrepancy between the results of
recent precise neutron lifetime measurements in magnetic \cite{Gonzalez2021}
and material \cite{Serebrov2018PhysRevC.97.055503} UCN traps. The
Monte-Carlo simulations may help to estimate UCN losses quite accurately, but a physical model
of diffuse neutron scattering by trap walls must be elaborated to raise more the
precision and reliability of these simulations.

\section{Acknowledgments}

The work of V.D.K. and P.D.G. is supported by the Russian Science Foundation
grant \# 23-22-00312.

\bibliographystyle{apsrev4-2}
%\bibliography{UCN}

\end{document}